\documentclass[times,authoryear,5p]{elsarticle}
\usepackage{graphicx,amssymb,color}

\journal{Icarus}
\begin{document}
\begin{frontmatter}
\title{Reassessing the origin of Triton}
\author{E. Nogueira}
\ead{erica.nogueira@on.br}
\address{Observatorio Nacional/MCT, Rua General Jos\'{e} Cristino 77; CEP 20921-400, Rio de Janeiro, RJ, Brasil}
\author{R. Brasser\corref{ca}}
\ead{brasser\_astro@yahoo.com}
\address{Dep. Cassiop\'{e}e, University of Nice - Sophia Antipolis, CNRS, Observatoire de la C\^{o}te d'Azur; F-06304, Nice, France}
\cortext[ca]{Corresponding author}
\author{R. Gomes}
\ead{rodney@on.br}
\address{Observatorio Nacional/MCT, Rua General Jos\'{e} Cristino 77; CEP 20921-400, Rio de Janeiro, RJ, Brasil}
\begin{abstract}
The origin of Neptune's large, circular but retrograde satellite Triton has remained largely unexplained. There is an apparent
consensus that its origin lies in it being captured, but until recently no successful capture mechanism has been found. Agnor
\& Hamilton (2006) demonstrated that the disruption of a trans-Neptunian binary object which had Triton as a member, and which
underwent a very close encounter with Neptune, was an effective mechanism to capture Triton while its former partner continued on a
hyperbolic orbit. The subsequent evolution of Triton's post-capture orbit to its current one could have proceeded through gravitational
tides (Correia, 2009), during which time Triton was most likely semi-molten (McKinnon, 1984). However, to date, no study has been
performed that considered both the capture and the subsequent tidal evolution. Thus it is attempted here with the use of numerical
simulations. The study by Agnor \& Hamilton (2006) is repeated in the framework of the Nice model (Tsiganis et al., 2005) to determine
the post-capture orbit of Triton. After capture Triton is then subjected to tidal evolution using the model of Mignard (1979, 1980).
The perturbations from the Sun and the figure of Neptune are included. The perturbations from the Sun acting on Triton just after its
capture cause it to spend a long time in its high-eccentricity phase, usually of the order of 10~Myr, while the typical time to
circularise to its current orbit is some 200~Myr, consistent with earlier studies. The current orbit of Triton is consistent with an
origin through binary capture and tidal evolution, even though the model prefers Triton to be closer to Neptune than it is today. The
probability of capturing Triton in this manner is approximately 0.7\%. Since the capture of Triton was at most a 50\%
event -- since only Neptune has one, but Uranus does not -- we deduce that in the primordial trans-Neptunian disc there were some 100
binaries with at least one Triton-sized member. Morbidelli et al. (2009) concludes there were some 1\,000 Triton-sized bodies in the
trans-Neptunian proto-planetary disc, so the primordial binary fraction with at least one Triton-sized member is 10\%. This value is
consistent with theoretical predictions, but at the low end. If Triton was captured at the same time as Neptune's irregular
satellites, the far majority of these, including Nereid, would be lost. This suggests either that Triton was captured on an orbit
with a small semi-major axis $a \lesssim 50$~$R_N$ (a rare event), or that it was captured before the dynamical instability of the Nice
model, or that some other mechanism was at play. The issue of keeping the irregular satellites remains unresolved.
\end{abstract}
\end{frontmatter}
\section{Introduction and background}
Of all the large natural satellites of the planets, Neptune's Triton is one of the most peculiar. It orbits Neptune at a distance of
354\,759~km or 14.3 Neptune radii ($R_N$), with a period of 5.877 days (Jacobson et al., 1991), similar to some of Uranus' regular
satellites. Its orbit is circular ($e \sim 10^{-5}$) (Jacobson et al., 1991), yet its inclination with respect to Neptune's
equator is 156.8$^\circ$, thus retrograde and ruling out a formation from within the Neptune system. Its retrograde orbit has led to
the belief that Triton was captured by Neptune from a heliocentric orbit (McKinnon, 1984; Benner \& McKinnon, 1995), and its
Neptune-centric orbit subsequently decayed through either tidal interaction (McCord, 1966; McKinnon, 1984; Goldreich et al., 1989;
Correia, 2009), Neptune's circumplanetary gas disc (McKinnon \& Leith, 1995) or a debris disc that formed from collisions among
Neptune's putative regular satellites (\'{C}uk \& Gladman, 2005). We shall briefly discuss each of these scenarios below. For an
excellent in-depth review article on Triton and its origin we refer to McKinnon et al. (1995).\\

\subsection{Capture}
Littleton (1936) hypothesised that Triton and Pluto originated as adjacent prograde satellites of Neptune and that ejection of the
latter left the former on a retrograde orbit. Yet McKinnon (1984) demonstrated that this scenario is impossible: the amount of mass
and angular momentum in the system is insufficient to make Triton retrograde. McKinnon (1984) argued instead that Triton and Pluto
originated from a reservoir of bodies in the outer solar system. Triton was captured by Neptune during a close approach and its orbit
was circularised through tidal evolution to its current one. McKinnon (1984) concluded that the tidal heating that followed as Triton
circularised should have melted it. However, McKinnon's (1984) capture scenario presented a problem. In order to capture Triton it
needs to lose enough energy to be bound from hyperbolic orbit and not be able to subsequently escape through the Hill sphere. This is
difficult to do in one orbit using tidal interaction alone. Therefore Benner \& McKinnon (1995) studied temporary capture from
heliocentric orbit and the subsequent evolution of a set of these temporary-captured orbits in the circular restricted three-body
problem consisting of the Sun, Neptune and a massless Triton. Solar perturbations acting on Triton cause its angular momentum to
oscillate with a period half of Neptune's orbital period, with secular perturbations acting on longer time scales. In extreme cases
these perturbations align and decrease Triton's pericentre distance, $q$, to within a few Neptune radii. Here the collision with an
existing satellite or aerodynamic drag from a putative circumplanetary nebula could have dissipated enough energy to make capture
permanent. However, Benner \& McKinnon (1995) favour prompt capture at low $q$ by collision or gas drag because a long-lasting
temporary capture would have resulted in large changes in $q$ and thus decreasing the probability of a collision or the effect of gas
drag. A prompt capture ensures there are many more close flybys which speed up the reduction of the orbit. In any case, permanent
capture was difficult to achieve.\\

A different scenario for the capture of Triton that did not require complex dynamics was proposed by Goldreich et al. (1989), who
favoured the idea of a collision having occurred between Triton and a hypothetical regular satellite of Neptune. Goldreich et al.
(1989) argued that a collision with a regular satellite containing a few percent of Triton's mass would have shattered the satellite
and left Triton bound to Neptune. However, this scenario could be problematic. The collision probability of Triton with a
regular satellite is $10^{-5}$ per pericentre passage. In order to make this scenario work with a reasonable probability, Goldreich et
al. (1989) argue there had to be 10\,000 Triton-mass objects encountering Neptune within 10~$R_N$, which is too large by several orders
of magnitude compared to recent estimates derived from the size-distributions of the Kuiper Belt and Jupiter's Trojans (Morbidelli et
al., 2009). However, it may not be necessary that each of these pericentre passages within 10~$R_N$ has to come from a unique
object. Instead, there just have to be of the order of 10$^5$ such passages of Triton-like objects. The question then becomes whether
this many passages is feasible. We shall return to this problem in the next section when we discuss our numerical simulations. Further
criticism of this scenario came from McKinnon et al. (1995), who argue that unless the satellite that Triton would collide with is
tiny and contains less than 2\% of the mass of Triton, Triton would have shattered too. Since a collision with the larger satellites is
much more likely, and assuming that Neptune's putative regular satellites were similar in size and mass to those of Uranus, neither the
satellite nor Triton would have survived the collision.\\

Given the difficulties of permanently capturing Triton in the above scenarios, Agnor \& Hamilton (2006) suggested a new idea. They
developed analytical arguments and used numerical simulations to show that Triton could be captured through the dissociation of a
binary planetesimal when it passed close to Neptune. These three-body encounters will disrupt the binary if its centre of mass
passes close enough to Neptune that the binary's orbital separation is approximately equal to its Hill sphere. Agnor \& Hamilton
(2006) tested this scenario with binaries consisting of objects with masses 1~$m_T$ and 0.1~$m_T$, where $m_T$ is the mass of Triton.
Agnor \& Hamilton (2006) showed that capture is plausible for a large variety of initial conditions of the binary, such as the velocity
at the time of encounter with Neptune and the separation of the binary. It turned out that the probability to capture Triton decreased
rapidly once the encounter velocity exceeded about 10\% of Neptune's orbital velocity, while the lighter body was captured for
encounter velocities of up to 70\% of Neptune's orbital velocity. This result is not surprising because the change in velocity
at disruption experienced by the heaviest body, $\Delta v_1$, is $\Delta v_1 \propto m_2$, while for the lighter body $\Delta v_2
\propto m_1$ (Agnor \& Hamilton, 2006), and thus the lighter body experiences the larger kick and is more easily captured. It then
automatically follows that for lighter binaries, while more easily disrupted because of their lower binding energy, it is also more
difficult to capture one of its members: the low binding energy translates into a very small $\Delta v$ and one of them is bound to
Neptune on a virtually parabolic orbit, where solar perturbations are stronger. Therefore, capture preceded by binary disruption is
favoured for heavy binaries or those with a large mass ratio. Agnor \& Hamilton (2006) found that the semi-major axis of Triton upon
capture would always exceed 300~$R_N$, while the lighter body could be captured much closer to Neptune. Unfortunately, Agnor \&
Hamilton (2006) do not provide any statistics for their mechanism so that the probability of this event having occurred could not be
tested nor compared to other results.\\

The lack of statistical information about the success rate of capturing Triton through the disruption of a planetesimal binary led
Vokrouhlick\'{y} et al. (2008) to investigate whether or not Triton and the other irregular satellites of the giant planets could be
captured via binary dissociation. This project was aimed at providing a comparison with the planet-planet encounter model of
Nesvorn\'{y} et al. (2007). The initial conditions of Vokrouhlick\'{y} et al. (2008) come from the Nice model (Tsiganis et al., 2005),
and they perform a series of simulations of the planetary instability, recording all the close encounters between the planets and
planetesimals. They subsequently send a very large number of binaries past each planet, with the distribution of the binary's
hyperbolic planetocentric orbits taken from the encounter parameters recorded earlier, and their size distribution taken from that of
Kuiper Belt objects (e.g. Bernstein et al., 2004). They had difficulty capturing many small satellites from the dissociation of
binaries, for reasons that we just explained above. When considering the capture of Triton they conclude that near-equal mass binaries
produce the most captures, with a mass ratio of 1:2 to 1:3 preferred.  While larger binary primary masses yield more captures, there
are fewer of these around so that the overall capture probability decreases. Together with the fact that capturing members from a
light binary is more difficult, this could explain why we see no sub-Triton mass irregular satellites orbiting Neptune and Uranus. In
conclusion, they find an overall capture probability of less than 2\%, with no restrictions on the semi-major axis, eccentricity or
inclination. They add that Triton's capture was most likely to have occurred 5-10~Myr after Neptune's formation when the planetesimal
disc was kept dynamically cold by the surrounding gas of the solar nebula; if it was captured during the planetary instability of the
Nice model, it is also preferred that it is captured early on before the disc is too dynamically hot and thus the encounter velocity
with Neptune is too large.\\

Thus, it appears that the favoured mechanism for the capture of Triton is through the dissociation of a binary that had a very close
encounter with Neptune. After capture, the orbit of Triton needs to shrink to its present size either through tides or other means.

\subsection{Post-capture evolution}
After Triton was captured and gravitationally bound to Neptune, several mechanisms have been invoked to evolve it to its current
orbit. These are tidal interaction with Neptune (McCord, 1966; McKinnon, 1984; Goldreich et al., 1989; Correia, 2009), interactions
with a circumplanetary gas disc (McKinnon \& Leith, 1995) or a debris disc (\'{C}uk \& Gladman, 2005). \\

The first attempt at calculating Triton's orbital history with tidal interaction was performed by McCord (1966), who used the tidal
model of MacDonald (1964) and expanded the equations expressing Triton's change in orbital elements up to sixth order in eccentricity.
By using reasonable estimates for Triton's Love number ($k_2$), quality factor ($Q$) and mass, McCord (1966) concludes that Triton
could have reached its current orbit from a highly extended, nearly parabolic orbit within the age of the solar system. The typical
time to become circular is some 100~Myr for $Q_T=100$, a typical value for rocky bodies. \\

The results of McCord (1966) were verified by McKinnon (1984) and Goldreich et al. (1989). McKinnon (1984) argues that the tidal
dissipation in Triton would have been enough to melt it, increasing the dissipation and thus shortening the time to reach its
current orbit. Goldreich et al. (1989), on the other hand, use a simplified tidal model valid only for eccentricity $e \ll 1$, but
because of angular momentum conservation they argue that for $e \sim 1$ the tidal evolution only changes the semi-major axis $a$ at
constant pericentre, $q$. Assuming a value $k_2/Q=10^{-3}$, typical for rocky bodies, they find that Triton could have reached its
current orbit from a post-capture orbit with $a=1\,000$~$R_N$ within 400~Myr, but argue that a semi-molten Triton would have increased
the rate of orbital decay. However, they added that the perturbations from the Sun cause repeated changes in the angular momentum with
a period equal to half of Neptune's orbital period. These perturbations in angular momentum cause $q$ to oscillate by as much as
10~$R_N$ when $a \gtrsim 600$~$R_N$. These oscillations greatly reduce the effect of tides and increase the time to circularise the
orbit, but the authors do not specify by how much.\\

Correia (2009) uses the tidal models of Mignard (1979, 1980) and Hut (1981), which do not require series expansions in the
eccentricity, and he includes keeping track of Triton's obliquity and spin rate. By assuming $k_2/Q_T = 10^{-3}$ and an initial spin
period of Triton of 24~h, Correia (2009) shows that Triton reaches its current orbit and spin rate in less than 1~Gyr from an orbit
with $a \sim 2\,000$~$R_N$ and $q\sim 7$~$R_N$. In addition, he argues that tides are sufficient to account for Triton's low
eccentricity ($e \sim 10^{-5}$) and obliquity ($\varepsilon_T=0^{\circ}.46$). Thus, from the above studies and with reasonable tidal
parameters, it appears as if tides raised on Triton by Neptune are capable of reducing it to its current orbit from a post-capture one
well within the age of the solar system, provided that Triton's pericentre upon capture is very close to Neptune ($q \sim 7$~$R_N$).\\

A different approach to reducing Triton's orbit from an extended post-capture one was performed by McKinnon \& Leith (1995), who
studied the influence of a circum-Neptunian gas disc on the orbit of a post-captured Triton. They mimic a putative nebula around
Neptune after a minimum-mass Uranus nebula. They report strong evolution of the eccentricity and semi-major axis but almost no change
in the inclination. Their results are insensitive to the radial surface distribution. Thus Triton could have evolved to its current
orbit through gas drag and subsequent tidal evolution because, after the gas has disappeared, Triton's eccentricity is about 0.2. Gas
drag is able to reduce Triton's orbital angular momentum to its current level in about 1\,000~years in the absence of solar
perturbations. The latter can increase the gas drag time scale to $10^4$--$10^5$~years, so that Triton could have outlived a hot,
turbulent nebula lasting some 1\,000~years, but not a cool, long-lived low-mass one ($10^6$~yr). In addition, McKinnon \& Leith (1995)
argue that Triton could have cleared an annulus in the gas that could have halted its orbital decay. Thus, even though gas drag seems a
favourable mechanism to circularise Triton from a post-capture orbit, it requires sensitive timing and the role of solar perturbations
might reduce its impact. Recently, Ayliffe \& Bate (2009) have performed the most sophisticated simulations of circumplanetary gas
discs and they do not encounter circumplanetary discs around protoplanets with masses similar to Uranus and Neptune. This may not be a
problem for Neptune because of the existence of Triton. However, Uranus' satellite system exhibits properties that are suggestive of a
disc origin. The latter was made popular by Canup \& Ward (2006) who discovered that the typical mass ratio of the regular
satellites relative to the giant planets should be $\sim 10^{-4}$. The Uranian satellites are all approximately an order of
magnitude less massive than that (Jacobson et al., 1992), suggesting a possible different origin than the one advocated by Canup \&
Ward (2006). One alternative is that the satellites formed through the viscous spreading of an impact-generated debris disc (Ward \&
Canup, 2003). In another alternative Ayliffe \& Bate (2009) suggest that a protoplanet's rotating envelope must cool following
the dispersal of the encompassing circumstellar disc. As it does so it may well flatten into a disc suitable for satellite growth,
suggesting a relatively late formation of the satellites. In any case, further study is needed to understand the formation
of the Uranian satellite system and the idea that Triton was captured by gas drag. \\

Both the tidal evolution model and the gas drag model have the disadvantage that they are most effective very close to Neptune.
However, it is likely that after capture Triton's semi-major axis was very large, $a \gtrsim 100$~$R_N$. Beyond this distance, solar
perturbations become important. Specifically, the Sun induces perturbations in Triton that are the same as the Kozai effect experienced
by high-inclination asteroids when perturbed by Jupiter (Kozai, 1962). The Kozai mechanism conserves the $z$-component of the angular
momentum, $l_z = \sqrt{1-e^2}\cos I$, where $I$ is the orbital inclination of Triton with respect to the orbit of the perturbing body
(the Sun, or Neptune's orbital plane). Hence for prograde orbits inclination and eccentricity oscillate out of phase, while for
retrograde orbits these oscillations are in phase. Depending on the initial conditions of the orbit, the Kozai effect can greatly
increase the pericentre distance of Triton after capture. In turn, this could greatly lengthen the time it takes for tides or gas drag
to circularise the orbit. These problems were pointed out by \'{C}uk \& Gladman (2005). They argued that the effect of the Kozai
oscillations of Triton's pericentre increases the time scale to reach a circular orbit beyond the age of the solar system and therefore
tides alone cannot circularise Triton. They reached their conclusion by averaging a simplified version of the tidal equations
over one Kozai cycle. In order to find a different and faster mechanism to circularise Triton they investigate the role of an
eccentric, retrograde Triton on Neptune's putative regular satellites. \'{C}uk \& Gladman (2005) argue that an eccentric Triton forces
a high eccentricity on Neptune's putative regular satellites, which begin to cross each other and collide within a few thousand years.
The collision would shatter both of these regular satellites and create a debris disc that Triton would pass through. By
modelling the effect of the disc as a series of impulsive kicks occurring at pericentre, they conclude that the time scale for the
evolution of Triton is $\sim 10^5$~yr, much shorter than for tides. The natural course of events is for Triton to sweep up all of the
mass in the disc through collisions. The added benefit of this fast circularisation is that it could save Nereid from being ejected or
colliding with Neptune. \\

Thus, the debris disc of \'{C}uk \& Gladman (2005) seems an interesting alternative to the tidal or gas drag models. Unfortunately
they do not specify in detail how they modelled the action of the debris disc on Triton so that its results cannot be verified. In
addition, their arguments might not hold because of the following. When considering Uranus' satellite system, perturbations from Triton
would increase their eccentricities to orbit-crossing values and their inclinations to a few degrees. Using the formulation of \"{O}pik
(1976), the collision probability between any of the regular satellites is also 10$^{-5}$ per orbit. Since the orbital period of the
regular satellites is much shorter than that of Triton just after its capture, these satellites should suffer several collisions before
any of them hits Triton. Most of these mutual collisions occur at impact velocities approximately 1 to 2 times the escape velocity of
the satellites. Agnor \& Asphaug (2004) show that collisions with such a low mutual velocity tend to be either merging or hit-and-run
collisions. For most of these hit-and-run collisions, the ejecta would be approximately 10\% of the mass of the smallest satellite
that is involved. Perturbations from the other satellites and Triton will quickly increase the eccentricities of the ejecta and
most of these are rapidly swept up. It is likely that eventually a collision with Triton will occur that will shatter both the
satellite and Triton. To summarise, it is not immediately clear whether these mutual collisions among the satellites will create the
debris disc that \'{C}uk \& Gladman (2005) suggested, or whether the satellites will remain largely intact and the system is destroyed
by a collision with Triton. \\

In addition to a possible collision, there is another outcome: ejection of Triton. The forcing of the eccentricities and inclinations
of the regular satellites by Triton depends on both the eccentricity of Triton and the semi-major axis ratio between Triton and the
satellites. For large semi-major axis ratios the eccentricities of the satellites could remain small enough to avoid crossing. Every
time Triton passes through the satellite system it receives a root-mean-square energy kick of the order of $\Delta (1/a) \sim Gm_T/a_s$
(Fern\'{a}ndez, 1981; Duncan et al., 1987), where $m_T$ is the mass of Triton and $a_s$ is the semi-major axis of a satellite. The
satellites receive an energy kick of the same magnitude. These kicks cause Triton to random walk in energy. Typically $a_s
\sim 10$~$R_N$ and so $\Delta (1/a) \sim 2 \times 10^{-5}$~$R_N^{-1}$. For a captured orbit similar to that of Nereid $1/a \sim 4
\times 10^{-3}$~$R_N^{-1}$ and the number of orbits to random walk to ejection is then around 40\,000, ignoring a possible rare
L\'{e}vy flight, and thus the time to eject Triton is some 10$^5$~years. These are crude estimates but what is important to notice
is that the ejection time and the collision time are very similar. Since most collisions leave the majority of the mass in the
satellites instead of debris, ejection becomes a feasible outcome. Clearly further study is needed to determine the most likely
scenario when considering a putative regular satellite system of Neptune being perturbed by a just-captured Triton.

\subsection{Current approach}
In this paper we investigate the capture and subsequent evolution of Triton in more detail, building on some of the earlier
works. First, we study the capture of Triton using the binary capture scenario of Agnor \& Hamilton (2006), but in the framework of
the Nice model, as in Vokrouhlick\'{y} et al. (2009). This model has booked a large number of successes, such as explaining the delay
that caused the Late Heavy Bombardment of the terrestrial planets (Gomes et al., 2005), the origin of Jupiter's Trojans (Morbidelli et
al., 2005), the structure of the Kuiper belt (Levison et al., 2008b), the dichotomy of the asteroid belt (Levison et al., 2009) and
the irregular satellites of the giant planets (Nesvorn\'{y} et al., 2007). Like Vokrouhlick\'{y} et al. (2009) we re-enact a series of
encounters with binary planetesimals. This will give us the distribution of orbits of Triton just after capture. Second, we use the
tidal equations of Correia (2009), based on the model of Mignard (1979, 1980), to determine which of the captured orbits are able to
circularise within the age of the solar system, and which initial conditions will place Triton approximately 14~$R_N$ away from Neptune
on a circular orbit. We improve upon Correia's (2009) tidal model, as well as those of McCord (1966) and Goldreich et al. (1989), by
adding the perturbations of the Kozai mechanism caused by the Sun, as suggested by \'{C}uk \& Gladman (2005), as well as the
perturbations from Neptune's figure. The latter should suppress the Kozai effect once the semi-major axis reaches $a_c \sim
73$~$R_N$ (Kinoshita \& Nakai, 1991). The final outcome of the capture and tidal simulations should give us a probability to obtain a
circular, retrograde Triton. This probability is then translated into the expected number of Triton-like objects in the proto-planetary
disc for consistency. This paper is divided as follows: in section~2 we present our models and employed methods. Section~3 contains the
results from the numerical experiments. In section~4 we discuss how Triton's capture affects Neptune's irregular satellites,
concentrating on Nereid. Conclusions and summary follow in the last section.

\section{Model and methods}
The model that we have employed consists of three parts. First, we record the encounters of planetesimals with Neptune during a
Nice model simulation. Second, the deepest encounters are re-enacted with binaries with varying mass ratio to determine the capture
probability and the resulting orbital distribution of captured objects. Third, the orbits of the captured orbits are evolved under the
action of tides to determine which ones end up being similar to Triton. Below we explain each stage in detail.

\subsection{Planet migration}
We consider a system consisting of the Sun, the four giant planets and a planetesimal disc. The initial solar system was more compact
than today, and the planets are thought to have formed between 5 and 15~AU on quasi-circular, coplanar orbits (Tsiganis et al., 2005).
We took the initial conditions of Gomes et al. (2005) which ensured that Jupiter and Saturn crossed their 2:1 orbital resonance some
600~Myr after their formation, which subsequently triggered the planetary instability (Gomes et al., 2005). We assumed that the
planetesimal disc was situated just beyond the orbits of the planets, ending at 30~AU, with radial mass distribution varying as
$r^{-1}$. The planetesimal disc was composed of $10\,000$ equal-mass bodies with a total mass equal to $35 M_\oplus$ (where $M_\oplus$
is the Earth's mass). We do not use the initial conditions that arise from the previous phase of migration induced by the gas disc
(Morbidelli et al., 2007) because we have a larger database of simulations of the former, and because the interactions between planets
and planetesimals are similar in both discs after the late instability. We simulated the dynamical evolution using the numerical
integrator MERCURY (Chambers, 1999), where we use a computational `trick' to decrease the amount of CPU time during the planetary
migration simulations (Gomes et al. 2004). We defined an encounter to occur when the distance between a planetesimal and a planet is
less than $d=f R_H$, where $R_H$ is the planetary Hill radius and $f$ is a factor larger than unity. We used $d=1$~AU. Each encounter
within this distance was registered in detail, keeping track of the position and velocity of the planetesimal and the planet in the
heliocentric reference frame, and the time of closest approach. Once a planetesimal entered the planet's Hill sphere, the system
switched to planetocentric coordinates, where it was observed that, as expected, most of the encounters are hyperbolic. However, some
of these encounters are elliptical and therefore the planetesimal remains around the planet for a relatively long time during a
temporary capture. The total integration time was 4.5~Gyr with a time step equal to 0.4~yr.\\

\begin{figure}
\resizebox{\hsize}{!}{\includegraphics[angle=-90]{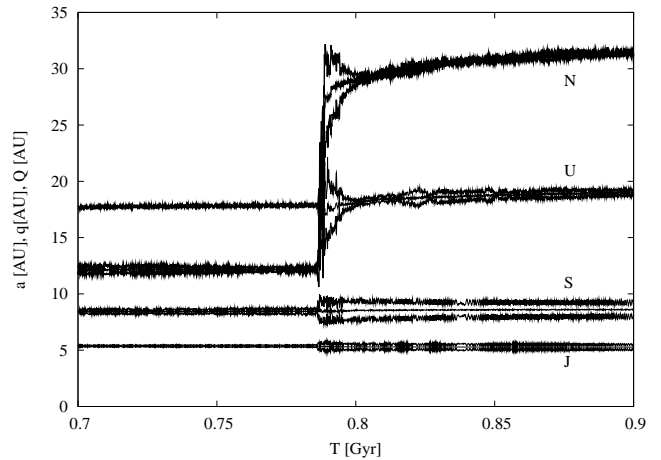}}
\caption{Semi-major axis, perihelion and aphelion distance of the four giant planets around the instability. The letters designate
which lines correspond to which planet. Neptune's path is traced with thick lines.}
\label{planets}
\end{figure}

We performed several Nice model simulations and chose the one where the final orbits of the giant planets are closest to their current
ones. Right from the beginning, some planetesimals from the disc encounter the outermost planet and consequently the planets
slowly migrate, increasing their orbital separation. This slow leakage from the disc continues until after 788~Myr when Jupiter
and Saturn cross the 2:1 resonance. This resonance crossing causes the system to destabilise and one of the ice giants (Neptune) is
scattered into the disc, destabilising it, and scattering planetesimals all over the solar system. In our preferred simulation, Saturn
and the ice giants undergo many mutual encounters and the ice giants exchange orbits. The number of planetesimals having close
encounters with the planets decreased quickly after the instability, and their population decays approximately exponentially. Some
50~Myr after the instability the planets are in their actual configuration. Fig.~\ref{planets} plots the evolution of the four giant
planets around the time of the instability. The lines represent their semi-major axis, perihelion and aphelion distance respectively.
The letters designate which set of lines corresponds to which planet: Jupiter stays around 5~AU, Saturn around 9~AU, Uranus goes from
18~AU to 20~AU and Neptune (outlined in thick lines) jumps from 12~AU to 30~AU. \\

\begin{table}
\begin{tabular}{ccccc}
Planet  & Number & Total enc & Elliptic & Hyperbolic \\ \hline \\
Jupiter & 5\,866 &  216\,921 &  8\,844 &  208\,077    \\
Saturn  & 8\,249 &  288\,213 &  4\,874 &  283\,339    \\
Uranus  & 9\,891 & 1\,375\,907 & 10\,997 & 1\,364\,910   \\
Neptune & 9\,934 & 1\,748\,867 & 65\,232 & 1\,683\,635
\end {tabular}
\caption{ First column: the total number of unique planetesimals that had close encounters with this planet. Second column: the total
number of close encounters between planetesimals and this planet. Third column: the total number of elliptical encounters between a
planetesimal and this planet. Fourth column: the total number of hyperbolic encounters between a planetesimal and this planet.}
\label{encst}
\end{table}

The encounter data from this preferred simulation is listed in Table~\ref{encst}. The first column lists the planet with which
the encounters occurred. The second column lists the number of unique test particles, out of 10\,000, that suffered an encounter with
this planet during the lifetime. The third column lists the total number of planetesimals that encountered the planet within 1~AU. The
fourth column lists the number of encounters that were elliptical while the last column lists the number of hyperbolic encounters.
There are several things that should be pointed out. The first is that Uranus and Neptune undergo many more encounters than Jupiter
and Saturn, because the former are scattered into the pristine planetesimal disc while Jupiter and Saturn are left behind. The
number of unique particles each planet encounters is similar to the ratios described in Fern\'{a}ndez (1997) based on planetesimals
that were scattered by Neptune from the Kuiper belt. Specifically, Jupiter typically ends up controlling 58\% of the planetesimals,
Saturn approximately 22\% (but it encounters 85\% of them) and Uranus and Neptune approximately 10\% each (so that each will
encounter about 90\% of the planetesimals). A second feature is that Neptune encounters substantially more planetesimals than Uranus,
partially because some particles that are under its control never encounter Uranus and instead suffer many encounters with Neptune on
their way to ejection, and partially because the proximity of Saturn to Uranus decreases the dynamical influence of the latter.
Vokrouhlick\'{y} et al. (2008) also reported a substantially lower number of planetesimal encounters with Uranus than with Neptune.
Last, the significantly larger number of elliptical encounters with Neptune vs Uranus is most likely the result of the higher encounter
velocities of the planetesimals with Uranus than with Neptune.\\

\begin{figure}
\resizebox{\hsize}{!}{\includegraphics[angle=-90]{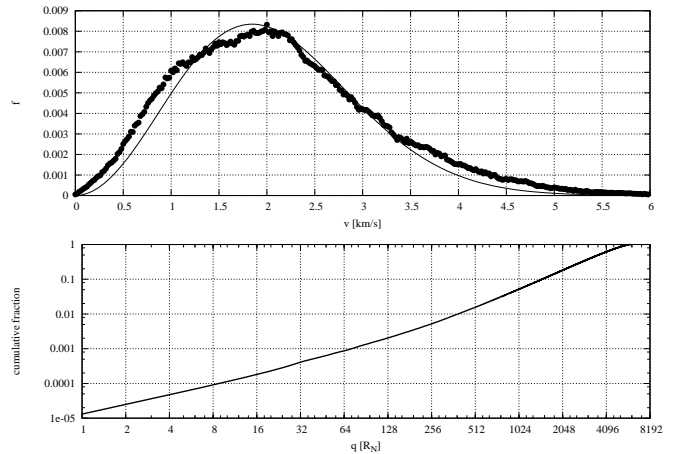}}
\caption{Top panel: velocity distribution of particles that encounter Neptune after the instability. The raw data is shown by the
bullets while the solid line shows the best-fit Maxwellian with $v_{\rm{m}}=1.31$~km~s$^{-1}$. Bottom panel: cumulative distribution in
pericentre distance with respect to Neptune.}
\label{encs}
\end{figure}

For our purpose we are only interested in planetesimal encounters with Neptune after the instability. For this study, we analyse
the distribution of the planetocentric orbits of the planetesimals that had close encounters with Neptune only. Specifically, we
are interested in the velocity distribution of the planetesimals as they encounter Neptune and the distribution of their closest
approach distance to Neptune. It turns out that the velocity distribution of the encounters is roughly Maxwellian, whose functional
form is given by 

\begin{equation}
 p(v) = \sqrt{\frac{2}{\pi}} \frac{v^2}{v_{\rm{m}}^3}\exp\Bigl(-\frac{v^2}{v_{\rm{m}}^2}\Bigr),
\end{equation}
with $v_{\rm{m}}$ the parameter velocity, which was found to be 1.31~km~s$^{-1}$. In the top panel of Fig.~\ref{encs} the
bullets depict the velocity distribution of the planetesimals as they encounter Neptune. The best-fit Maxwellian is plotted as a solid
line. The bottom panel of Fig.~\ref{encs} plots the cumulative distribution of the peri-Neptune distance, $q$. For small distances the
cumulative distribution is linear in $q$ while beyond 200~$R_N$ the fit scales as $q^2$, suggesting that gravitational focusing is only
important for very close encounters. Both plots are in good agreement with those presented in Vokrouhlick\'{y} et al. (2009). \\

The cumulative $q$ distribution in the bottom panel of Fig.~\ref{encs} can be used to constrain the collisional capture scenario
of Goldreich et al. (1989), which was mentioned in the introduction. Given that the probability of Triton colliding with a fictitious
Neptunian regular satellite is approximately 10$^{-5}$ for each peri-Neptune passage, one needs of the order of 10$^5$ passages of
Triton-like objects within approximately 20~$R_N$ in order to capture it through a collision event. From the bottom panel of
Fig.~\ref{encs} the probability per encounter of a planetesimal coming within 20~$R_N$ is of the order of $2 \times 10^{-4}$. Combined
with the intrinsic collision probability with these fictitious satellites Neptune needs to undergo of the order of $10^9$ encounters
with Triton-like objects. Using the data from Table~\ref{encst} as guidance, on average each planetesimal undegoes about 100 encounters
with Neptune, so that for each Triton-like object the probability of collisional capture is approximately 10$^{-7}$. This low
probability makes the collisional capture scenario seem a very unlikely mechanism to account for the existence of Triton.\\

Now that we have a data base of encounters with Neptune, the next step is to re-enact these encounters using binaries and
determine which binary systems are disrupted and also leave Triton bound to Neptune.

\subsection{Binary encounters}
After recording the number of encounters with Neptune from the migration simulation, we re-enacted these using binary systems. The
centre of mass of the binary coincides with the trajectory of a planetesimal that had a close encounter with Neptune. From Agnor \&
Hamilton (2006) we know that a binary is ionised once the separation between the members becomes equal to the Hill sphere of the
binary. This tidal disruption distance is given by

\begin{equation}
\label{rd}
\frac{r_{td}}{R_N}=\left(\frac{a_{B}}{R_{1}}\right)\left[\left(\frac{3\rho_{N}}{\rho_{1}}
\right)\left(\frac{m_{1}}{m_{1}+m_{2}}\right)\right]^{1/3} \approx \frac{a_B}{R_1},
\end{equation}
where $R_N$ is Neptune's radius, $R_1$ is the radius of the primary binary component (usually Triton's radius), $\rho_N$ is the mean
density of Neptune, $\rho_1$ is the mean density of the primary component of the binary (usually Triton's density) and $m_1$ and $m_2$
are the masses of the primary and secondary of the binary. The approximation sign in equation~(\ref{rd}) above is valid when
considering densities appropriate for Neptune and Triton. We created four different groups of binaries with different mass ratios,
with each binary containing one member with a mass equal to Triton's mass. The other member was either 0.1~$m_T$, 0.3~$m_T$, 1~$m_T$ or
3~$m_T$. All binaries had an initially circular orbit and semi-major axis $a_B = 1.5$~$R_N = 37\,146$~km, which is the approximate
maximum value for the currently-known trans-Neptunian binary population (Noll et al., 2008). These binaries should disrupt
once they come closer to Neptune than $q \sim 27$~$R_N$. The other orbital elements were chosen at random. For each binary group we
simulated 1\,000 different orbits. To be on the safe side, we re-enacted only encounters with $q<100$~$R_N$. \\

After we created the binary system, we simulated their approach with Neptune using the MERCURY integrator, and registered the changes
in semi-major axis, eccentricity and inclination of the binary. We consider that a member is ejected from the system whenever it has
$a>3000$~$R_N$. If the pericentre distance is smaller than the planetary radius the body is removed from the simulation.\\

This part of the model allowed us to create a distribution of orbits that Triton could have had just after capture. The next step is
to evolve these orbits using tides raised by Neptune on Triton to determine what fraction of post-capture orbits will yield the
current orbit of Triton within the age of the solar system.

\subsection{Tidal evolution}
\begin{figure}
\resizebox{\hsize}{!}{\includegraphics[angle=-90]{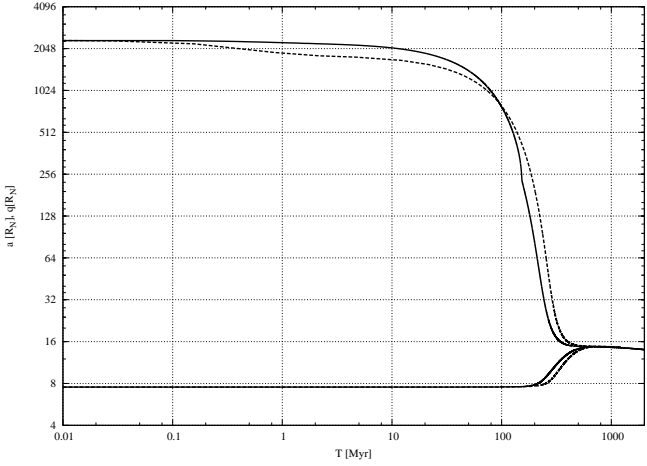}}
\caption{Tidal evolution of semi-major axis and pericentre distance of a Triton-sized object captured by Neptune. Initial conditions
are from Correia (2009). Solid lines depict the evolution from Correia (2009) while the dashed lines show our simplified
implementation. Solid lines data set courtesy of Alexandre Correia.}
\label{tidesexample}
\end{figure}

We decided to use the tidal equations presented in Correia (2009), which are based on the tidal models of Mignard (1979, 1980) and Hut
(1981). This model assumes that the time delay of the tidal response of a body, $\Delta t$, is a constant. This model can be made
linear in $\Delta t$ and the tidal equations can be written in closed form for all values of eccentricity $e< 1$. However, the
constant time delay model does not appear to agree with geophysical data (Efroimsky \& Lainey, 2007). The rate of energy
dissipation within a body is characterised by the tidal parameter $Q$, which is a measure of how many tidal oscillations are needed to
damp the energy by order of itself. The constant time delay model of Mignard (1979, 1980) assumes that $Q$ is inversely proportional to
the frequency with which the body is distorted, $\chi$. For eccentric orbits $\chi$ is the angular frequency of this body at
pericentre. Laboratory experiments indicate that $Q \propto \chi^{\alpha}$, where $\alpha \sim 1$ for the range of periods that we are
interested in (Karato, 2008). Nevertheless, we are mostly dealing with orbits with $e \sim 1$ where series expansions are invalid, and
thus we decided to adopt this model because it best suits our needs. We decided to use a simplified model in which we integrate the
equations governing the changes in semi-major axis, eccentricity, spin rate and obliquity of Triton, while ignoring any changes in the
rotation rate of Neptune and its obliquity. This is justified because Correia (2009) found that the change in both of these quantities
is negligible. We found that tides raised by Triton on Neptune are a couple of orders of magnitude weaker than tides raised by Neptune
on Triton, but we included them for the sake of completeness. The tidal equations are given by (Correia, 2009)

\begin{eqnarray}
 \dot{a}_T &=& \frac{2K_T}{m_Ta_T}\Bigl(\frac{f_2(e_T)\cos \varepsilon_T \omega_T}{n_T}-f_3(e_T)\Bigr) \nonumber \\
&+&\frac{2K_N}{m_Ta_T}\Bigl(\frac{f_2(e_T)\cos i_T \omega_N}{n_T}-f_3(e_T)\Bigr) \nonumber \\
\dot{e}_T &=& \frac{9K_Te_T}{m_Ta_T^2}\Bigl(\frac{11 f_4(e_T)\cos \varepsilon_T \omega_T}{18 n_T}-f_5(e_T)\Bigr) \nonumber \\
&+&\frac{9K_Ne_T}{m_Ta_T^2}\Bigl(\frac{11 f_4(e_T)\cos i_T \omega_N}{18 n_T}-f_5(e_T)\Bigr), \nonumber \\
\dot{\omega}_T &=&-\frac{K_Tn_T}{C_T}\Bigl(f_1(e_)\frac{1+\cos^2 \varepsilon_T}{2}\frac{\omega_T}{n_T}-f_2(e_T)\cos
\varepsilon_T\Bigr), \nonumber \\
\dot{\varepsilon}_T &=&\frac{K_Tn_T}{C_T\omega_T}\sin \varepsilon_T \Bigl(\frac{f_1(e_T)\cos \varepsilon_T
\omega_T}{2n_T}-f_2(e_T)\Bigr),
\label{tideseq}
\end{eqnarray}
where $\varepsilon_T$ is Triton's obliquity with respect to its own orbit, $i_T$ Triton's inclination with respect to Neptune's
equator, $\omega_T$ is Triton's spin rate, $\omega_N$ is Neptune's spin rate, $n_T$ is Triton's mean motion, $C_T$ is Triton's moment
of inertia along its spin axis, and

\begin{eqnarray}
K_T&=&\frac{3k_{2T}Gm_N^2R_T^5\Delta t_T}{a_T^6}, \\
K_N&=&\frac{3k_{2N}Gm_T^2R_N^5\Delta t_N}{a_T^6}, \nonumber \\
f_1(e) &=& (1+3e^2+3e^4/8)(1-e^2)^{-9/2}, \nonumber \\
f_2(e)&=&(1+15e^2/2+45e^4/8+5e^6/16)(1-e^2)^{-6}, \nonumber \\
f_3(e)&=&(1+31e^2/2+255e^4/8+185e^6/16+25e^8/64) \nonumber \\
&\times&(1-e^2)^{-15/2} \nonumber \\
f_4(e)&=&(1+3e^2/2+e^4/8)(1-e^2)^{-5}, \nonumber \\
f_5(e)&=& (1+15e^2/4+15e^4/8+5e^6/64)(1-e^2)^{-13/2}. \nonumber 
\label{kandf}
\end{eqnarray}
Here $k_{2T}$ is Triton's Love number $k_2$ and $\Delta t_T$ is Triton's tidal response time, $R_T$ is Triton's radius and $m_T$ is
its mass. The quantities carrying subscripts N are for Neptune. Table~\ref{quant} lists the values of the various
quantities used above. Most are taken from Correia (2009), and references therein. The values of $\omega_T$ and $\varepsilon_T$ are
starting values, which evolve to their current ones. An example of the tidal evolution in this simplified model is depicted in
Fig.~\ref{tidesexample}, and shows the evolution of semi-major axis and pericentre distance of a Triton-sized object captured by
Neptune. The initial conditions are taken from Correia (2009) and are $a_0=2354$~$R_N$, $q_0=7$~$R_N$, $i_0=157^\circ$ and
$\omega_{r0} = 1/24$~hr$^{-1}$. The solid lines show the evolution depicted in Correia (2009) while the dashed lines show the
evolution according to our simplified model. The slight differences are caused by Correia (2009) taking Cassini states and 
properly taking Triton's obliquity evolution into account.\\

We considered two values of $\Delta t_T$. The first, $\Delta t_T = (Q_T\omega_T)^{-1}= 808$~s, where we used the {\it current}
value of $\omega_T$ and corresponds to imposing $Q_T=100$, a commonly adopted value for rocky bodies and close to the inferred value
for Mars of $Q_M=92$ (Yoder et al., 2003). However, Triton was most likely semi-molten during its tidal circularisation (McKinnon,
1984), which decreases the value of $Q$ and thus increases that of $\Delta t_T$. Using Io as an example, it has $k_2/Q_I \approx 0.015$
and thus $Q_I \sim 3$ for $k_2=0.05$ (Lainey et al., 2009) but when using the fluid Love number, appropriate for semi-molten bodies,
$k_2=1.292$ (Anderson et al., 2001) and we have $Q_I \sim 86$. For solid bodies, we can turn to the Moon and Mars. For the Moon, the
inferred value of $k_2$ from laser ranging is $k_2=0.02664$ and $Q \sim 30$ (Williams et al., 2005). This low value of $Q$ is caused by
tidal dissipation within the Moon's core, and results in $k_2/Q \sim 0.0011$. For Mars, its value of $k_2$ ranges from 0.11 to 0.16
(Marty et al., 2009) while its value of $Q$ ranges between 80 (Lainey et al., 2007) and 92 (Yoder et al., 2003), suggesting that for
Mars $k_2/Q \sim 0.0015$. Thus, it seems that for rocky bodies $k_2/Q \sim 0.0013$ while for semi-molten bodies $k_2/Q \sim 0.015$. We
decided to adopt an intermediate value of $k_2=0.1$ for Triton (McKinnon et al., 1995) and change the value of $Q_T$ by a factor ten
between the solid and semi-molten state. For the solid state we adopt $k_2/Q_T = 10^{-3}$, implying $Q_T=100$, and for the semi-molten
state $k_2/Q_T=10^{-2}$, implying $Q_T=10$. This corresponds to a time delay $\Delta t = 808$~s in its current orbit for the solid
state, and 8080~s when semi-molten.\\

\begin{table}
\begin{tabular}{cc}
Quantity & Value \\ \hline \\
$m_T$ & $1.0837 \times 10^{-8}$~M$_{\odot}$ \\
$m_N$ & $5.156 \times 10^{-5}$~M$_{\odot}$ \\
$k_{2T}$ & 0.1 \\
$k_{2N}$ & 0.407 \\
$R_T$ & $9.044 \times 10^{-6}$~AU (1353~km)\\
$R_N$ & $1.6554 \times 10^{-4}$~AU (24764~km) \\
$C_T$ & 0.35$m_TR_T^2$ \\
$\omega_T$ & 6884.65 rad~yr$^{-1}$ ($P_T=8$~h) \\
$\omega_N$ & 3418.82 rad~yr$^{-1}$ ($P_N = 16.11$~h) \\
$\varepsilon_T$ & 170$^\circ$ \\
$\varepsilon_N$ & 28.56$^\circ$ \\
$\Delta t_N$ & 1.02~s ($Q_N = 9000$) \\
$\Delta t_T$ & 808~s or 8080~s \\
$a_{\odot}$ & 30.1~AU \\
$e_{\odot}$ & 0.008 \\
$J_2$ & $3.343 \times 10^{-3}$
\end{tabular}
\caption{Values of various quantities that enter the tidal equations, as well as those of the Kozai mechanism and figure of
Neptune.}
\label{quant}
\end{table}

As stated in the introduction, the tides are not acting alone. For large orbits, the Kozai mechanism (Kozai, 1962) is at work, caused
by perturbations from the Sun, and we need to take its influence into account (\'{C}uk \& Gladman, 2005). The Kozai mechanism has two
constants of motion (Kozai, 1962; Kinoshita \& Nakai, 2007)

\begin{eqnarray}
H_K&=&\gamma[(2+3e^2)(3\cos^2 I -1)+15e^2\sin^2 I \cos 2\omega], \nonumber \\
h_z &=& (1-e^2)^{1/2}\cos I,
\end{eqnarray}
where we omit the subscript T since we are only dealing with Triton. Here $H_K$ is the averaged Kozai Hamiltonian (Kozai, 1962), $h_z$
is the $z$-component of the orbital angular momentum, $I$ is Triton's inclination with respect to Neptune's orbital plane, $\omega$ is
Triton's argument of pericentre and

\begin{equation}
 \gamma = \frac{n_{\odot}^2a_T^2}{16(1-e_{\odot}^2)^{3/2}}.
\end{equation}
The Kozai mechanism induces coupled oscillations in eccentricity and inclination, and either a circulation or libration of $\omega$,
depending on the value of $H_K$ and $h_z$. The equations of motion of these three variables, and $\Omega$, the longitude of Triton's
ascending node on Neptune's orbital plane (see below) are (Kinoshita \& Nakai, 2007)

\begin{eqnarray}
\dot{e}&=&\frac{30\gamma}{na^2}e(1-e^2)^{1/2}\sin^2 I \sin 2\omega, \nonumber \\
\dot{I}&=& -\frac{15\gamma}{na^2(1-e^2)^{1/2}}e^2\sin 2I \sin 2\omega, \nonumber \\
\dot{\omega}&=&\frac{6\gamma}{na^2(1-e^2)^{1/2}}[-1+e^2+5\cos^2 I \nonumber \\
&+&5(1-e^2-\cos^2 I)\cos 2\omega], \nonumber \\
\dot{\Omega}&=&-\frac{6\gamma}{na^2(1-e^2)^{1/2}}\cos I(3e^2+2-5e^2\cos 2\omega).
\label{kozaieq}
\end{eqnarray}

In addition to the Kozai mechanism, the figure of Neptune perturbs Triton's orbit once it is close enough to Neptune. While the
perturbations of Neptune's figure do not alter the eccentricity and inclination, it does force a precession of the argument of
pericentre. Once Triton is close enough to Neptune, this precession will overtake the precession induced by the Kozai effect and the
coupled oscillations in inclination and eccentricity will stop. Thus, in order to model the disappearance of the Kozai mechanism, we
need to add the perturbations of Neptune's figure to the tidal model as well. The averaged Hamiltonian is (e.g. Kinoshita \& Nakai,
1991)

\begin{equation}
 H_{J2} = \frac{1}{4}\frac{Gm_N}{a}J_2\Bigl(\frac{R_N}{a}\Bigr)^2\frac{(3\cos^2 i -1)}{(1-e^2)^{3/2}},
\end{equation}
where $i$ is Triton's inclination with respect to Neptune's equator and $J_2$ is Neptune's quadrupole moment. In order to add this to
the tidal equations of motion and those of the Kozai mechanism, we need to transform Triton's inclination with respect to Neptune's
equator to its inclination with respect to Neptune's orbital plane. This is done via $\cos i = \cos \varepsilon_N \cos I - \sin
\varepsilon_N \sin I \cos \Omega$, where $\varepsilon_N$ is Neptune's obliquity. Unfortunately this adds another variable, $\Omega$, to
be integrated, and therefore the Kozai part of the regression of this angle needs to be taken into account as well (hence its
inclusion above). We have

\begin{eqnarray}
\dot{I}&=&-\frac{3}{2}\frac{Gm_N}{a}\frac{J_2}{na^2(1-e^2)^2}\Bigl(\frac{R_N}{a}\Bigr)^2\cos i \sin \varepsilon_N \sin \Omega,
\nonumber \\
\dot{\omega}&=& \frac{3}{4}\frac{Gm_N}{a}\frac{J_2}{na^2(1-e^2)^2}\Bigl(\frac{R_N}{a}\Bigr)^2 \nonumber \\
&\times&\Bigl[\cot I \sin
2i\frac{di}{dI}+3\cos^2i-1\Bigr], \nonumber \\
\dot{\Omega} &=& -\frac{3}{4}\frac{Gm_N}{a}J_2\Bigl(\frac{R_N}{a}\Bigr)^2\frac{\csc I \sin 2i}{na^2(1-e^2)^2}\frac{d i}{dI}.
\label{j2eq}
\end{eqnarray}

We integrated the system of equations consisting of (\ref{tideseq}), (\ref{kozaieq}) and (\ref{j2eq}) using a Bulirsch-Stoer
integrator with variable time step (Bulirsch \& Stoer, 1966). Once Triton was closer than 20~$R_N$ to Neptune, we stopped
integrating the Kozai and $J_2$ effects because the former is suppressed by the $J_2$ precession and the latter only causes
circulation of $\Omega$ and $\omega$, which do not contribute to the tidal evolution. Switching off the integration of these quantities
significantly sped up the final part of the tidal evolution. \\

When considering pure tidal motion, the orbital angular momentum of Triton is approximately conserved. Thus, for initially very
eccentric orbits, the final, circular, orbit is located at $a \approx 2q_0$, where $q_0$ is the original pericentre distance.  Triton's
current orbit suggests it was captured with $q_0 \sim 7$~$R_N$. However, the Kozai mechanism induces oscillations in the eccentricity
and thus in $q$. From equations (\ref{tideseq}) we have $\dot{a} \propto q^{-15/2} + O(q^{-13/2})$, so that a typical time scale on
which the tides act, $T_a = a/\dot{a} \propto q^{15/2} +O(q^{13/2})$ is a very steep function of $q$ and suggests the annulus in which
the tides can circularise Triton within the age of the solar system is very narrow. Indeed, increasing $T_a$ by an order of magnitude
requires a relative increase in $q$ of only 35\%. Experimentation showed that the time to become circular reaches the age of the
solar system for $Q_T=10$ when $q_0=20$~$R_N$, and thus orbits for which $q$ never dips below 20~$R_N$ can be ignored. Given the rapid
increase in $T_a$ with $q$, in the simplest and crudest sense one can envision the $q$ dependence as a step function, where the tides
are switched off if $q$ is larger than some threshold value, and the tides are active when $q$ is smaller. This behaviour suggests that
the final orbit of Triton would have $a \approx 2q_{\rm min}$, where $q_{\rm min}$ is the minimum value of $q$ that is obtained during
the Kozai cycle. Indeed this appears to be a better approximation than the previous one $a \approx 2q_0$.\\

Now that we have all the ingredients in place, we report the results of our experiments below. In what follows, we removed any objects
that achieved a final semi-major axis $a<5$~$R_N$ since it would then collide with Proteus (located at 4.8~$R_N$), or which encountered
$q<1$~$R_N$ during their Kozai cycle, or whose $q$ never reached below 25~$R_N$ within 1~Myr (by comparison: the Kozai cycle time is
typically 0.01~Myr). While we did integrate cases with $Q_T=100$, we do not present the results here since it was most likely that
Triton was semi-molten (McKinnon, 1984). In what follows, the time scale for the tidal evolution of Triton should be considered as
indicative rather than absolute. This is caused by the lack of knowledge of the tidal parameters and the model's simplification of the
real tidal evolution.

\section{Results}
In this section the results from our numerical simulations are presented. First we present the distribution of captured orbits.
This is followed by a case study of the tidal evolution, after which we turn to the tidal evolution of all the captured orbits.

\begin{figure}
\resizebox{\hsize}{!}{\includegraphics[angle=-90]{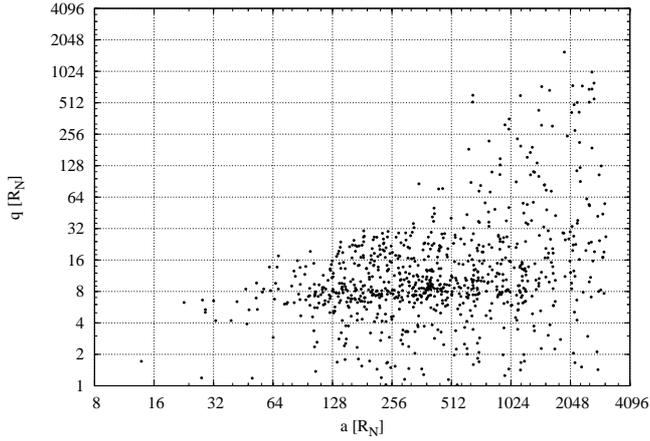}}
\caption{Pericentre distance vs. semi-major axis of captured satellites with Triton's mass after the disruption of a binary. These
are orbits only upon capture without subsequent tidal evolution.}
\label{captorb}
\end{figure}

\subsection{Captured orbits}
\begin{figure}
\resizebox{\hsize}{!}{\includegraphics[angle=-90]{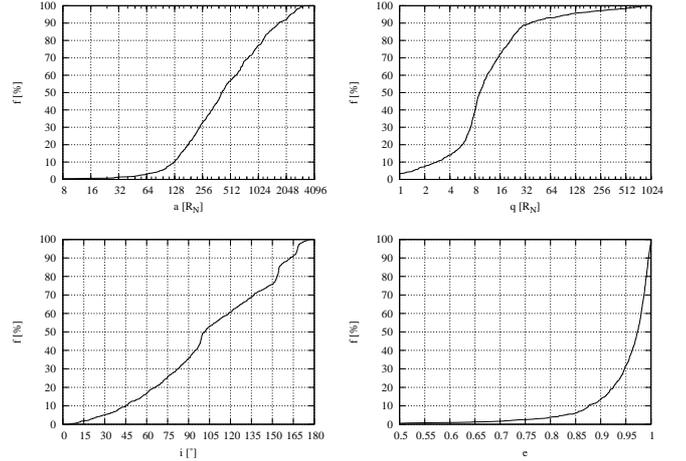}}
\caption{Cumulative distributions of semi-major axis (top-left), pericentre (top-right), inclination (bottom-left) and eccentricity
(bottom-right) for Triton-mass satellites upon capture.}
\label{captcum}
\end{figure}

Fig.~\ref{captorb} displays the pericentre distance ($q$) vs. semi-major axis ($a$) of objects with Triton's mass that were
successfully captured following the disruption of a binary, as outlined in Section 2.2. Since the binaries were enacted from
encounters with Neptune that had $q < 100$~$R_N$, it is no surprise that most captured objects have a rather small value of $q$. The
truncation of the semi-major axis at approximately 3000~$R_N$ is caused by our condition that objects with $a>3000$~$R_N$ are
considered unbound since they will be outside Neptune's Hill sphere at apocentre. Fig.~\ref{captcum} shows the cumulative distributions
of semi-major axis (top-left), pericentre (top-right), inclination (bottom-left) and eccentricity (bottom-right) for Triton-mass
satellites upon capture. The steep increase in the distribution of $q$ between 8~$R_N$ and 32~$R_N$ is probably caused by our choice of
binaries. Similarly, the sharp rise in the eccentricity distribution when $e \gtrsim 0.9$ is might also be an artefact of our initial
conditions. The inclination distribution shows there is a clear preference for retrograde orbits upon capture, probably
because of the increased stability of retrograde orbits with respect to the size of Neptune's Hill sphere (Hamilton \& Krivov, 1997).
In summary, most orbits are captured with $a \in (100, 2000)$~$R_N$ and $q \in (4,32)$~$R_N$ with a clear retrograde preference.
In contrast, Agnor \& Hamilton (2006) find that the median semi-major axis of Triton after capture is $a \gtrsim 1\,000$~$R_N$, while
we find a value of $a \sim 500$~$R_N$, suggesting that the encounter parameters when Neptune migrates are different from the static
case. These orbits now need to be evolved using the tidal model presented above. We define a successful case when Triton achieves
$e \sim 10^{-5}$ in less than 4~Gyr. In principle one could define a successful case as any capture for which $e <1$ and
$q > 1$~$R_N$ but then the current orbit is not reproduced and thus we decided to use the former criterion.

\subsection{Tidal evolution: Case study}
In order to determine whether or not we can obtain Triton's current orbit within the age of the solar system from the captured orbits
presented in the previous subsection, we need to run them through the tidal equations. In this subsection we show a representative
case of the evolution of Triton upon capture. In the next subsection we present the results of the final orbits of all captured
objects. We have decided to model the tidal evolution using two different methods, in order to compare them. For the first method, we
integrate the tidal equations from section 2.2 using as the initial conditions the captured orbits presented above. We included Kozai
and Neptune's figure and set Neptune's obliquity equal to its current value. With this method one can obtain Triton's final
inclination with respect to Neptune's current equator. For the second case, we use the same equations as above but set Neptune's
obliquity equal to zero. This was done to determine how the Kozai perturbations from the Sun determine the final outcome. \\

\begin{figure}
\resizebox{\hsize}{!}{\includegraphics[angle=-90]{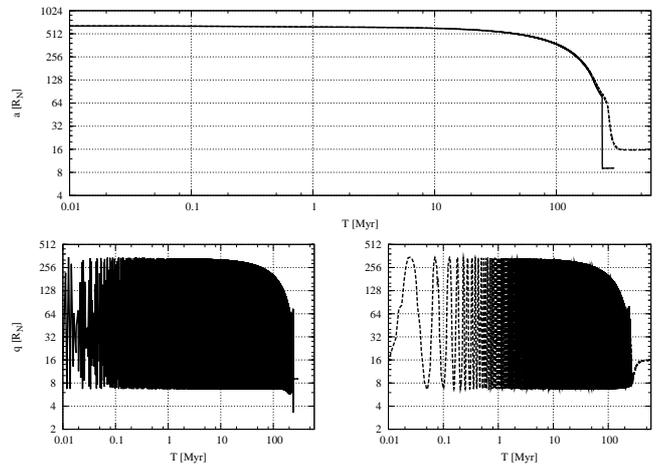}}
\caption{Example of tidal evolution of one object using two different models. The top panel shows the evolution in semi-major axis.
The solid line uses Neptune's current obliquity. The dashed line has Neptune's obliquity set to zero. The bottom panels show the
evolution of the pericentre. Left panel has the current obliquity of Neptune while the right panel has Neptune's obliquity set to
zero.}
\label{tides}
\end{figure}

Fig.~\ref{tides} presents the results of the tidal evolution of one object using $Q_T=10$ i.e. the semi-molten case. The starting
conditions are $a=648.2$~$R_N$, $q=8.38$~$R_N$, $I=52.1^\circ$, $\omega=207.1^\circ$. The top panel shows the evolution of the
semi-major axis vs. time. The solid line depicts the case with Neptune's current obliquity $\varepsilon_N = 28.56^\circ$ (case
1). The dashed line depicts the case where the obliquity of Neptune is zero (case 2). In both cases the pericentre distance oscillates
with large amplitude and short period compared to the tidal evolution. The latter occurs quickest when $q$ is at a minimum, and when
this happens the semi-major axis shrinks. With each decrease in semi-major axis at minimum $q$, the value of $h_z$ decreases and the
excursions in $q$ (and $e$) are less pronounced. Eventually the semi-major axis shrinks to the point where the perturbations from
Neptune's figure compete with the Kozai cycles induced by the Sun. This occurs at approximately 70~$R_N$. Once the semi-major
axis shrinks below this value the Kozai cycles cease and the rest of the evolution proceeds at constant angular momentum. However,
this value of the angular momentum is not equal to the value upon capture. The final semi-major axis is approximately $a_{\rm{f}} \sim
2q_{\rm{min}}$.\\

\begin{figure}
\resizebox{\hsize}{!}{\includegraphics[angle=-90]{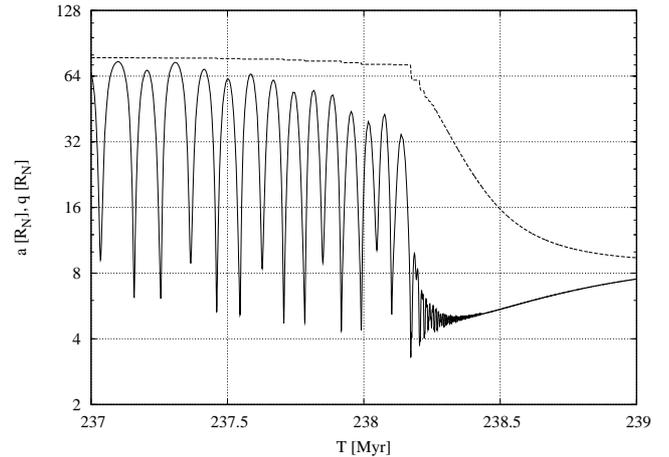}}
\caption{Zoom of the tidal evolution of Fig.~\ref{tides} around the time the Kozai mechanism stops. The dashed line shows the
semi-major axis vs time while the solid line shows the pericentre distance. The minimum $q$ decreases, due to a misalignment between
Neptune's rotational pole and orbital pole, until eventually it reaches below 4~$R_N$. The tidal evolution is then so rapid that the
semi-major axis shrinks sufficiently for the Kozai cycles to stop. The remaining evolution proceeds at constant angular momentum.}
\label{tideszoom}
\end{figure}

Fig.~\ref{tideszoom} presents a zoom for case 1 around the time when the Kozai cycles stop. The dotted line plots the semi-major axis
vs time while the solid line represents the pericentre distance, $q$. The oscillations in $q$ decrease in amplitude and the minima are
consistently closer to Neptune. The decrease in the minimum distance of $q$ is caused by the perturbations from Neptune's figure not
being aligned with those of the Sun. Far from Neptune the orbital angular momentum vector of Triton precesses perpendicular to
Neptune's orbital plane. As the semi-major axis of Triton shrinks, the perturbations from Neptune's figure become stronger compared
to the Kozai cycles induced by the Sun. Neptune's equator is not aligned with its orbit and the perturbations from Neptune's figure
force Triton's orbital angular momentum to precess perpendicular to Neptune's equator rather than its orbit. The transition happens
around 70~$R_N$. As the angular momentum gradually starts to precess about Neptune's rotational pole, the value of $h_z$, which is a
constant for the Kozai motion, has to decrease, causing an increase in the maximum eccentricity and thus a decrease in the minimum
value of $q$. Eventually the minimum $q$ has dropped close to 4~$R_N$ and the tidal evolution is then so rapid that the semi-major axis
shrinks considerably before Kozai cycles lift the pericentre again. However, by this time, the amplitude of the Kozai cycles
have decreased considerably so that future cycles are quickly damped and the semi-major axis continues to decrease. By now the tidal
evolution more or less conserves the current total angular momentum since the Kozai cycles have stopped, and the orbit circularises at
$a \sim 2q_{\rm{min}}$. \\

The outcome for the two simulations are different and are presented in Table~\ref{tidesoutcome}. The first column represents the
elements $a$, $q$ and time to become circular ($T_c$). The second column lists the initial values. The third column marks the first
case (current obliquity of Neptune). The fourth column marks the second case (no obliquity of Neptune). As one can see, the first case
yields an orbit much closer to Neptune than the second case. The time to become circular varies by a factor of two. The difference
between the final semi-major axis and values of $T_c$ are the result of the difference in Neptune's obliquity. For the first case the
orbital evolution proceeds very quickly once $q \sim 4$~$R_N$, which does not occur in the second case, where $q$ never drops below
$\sim 7$~$R_N$.\\

\begin{table}
\begin{tabular}{cccc}
Element & Initial & 1 & 2\\ \hline \\
$a$ & 648.2 & 9.03 & 15.68 \\
$q$ & 8.38 & 9.03 & 15.68 \\
$T_c$ & 0 & 242.3 & 631.8
\end {tabular}
\caption{The initial and final semi-major axis ($a$), pericentre ($q$) and time to circularise ($T_c$) for the sample orbit with two
different tidal models. The column (1) depicts the fist case where Neptune has its current obliquity. The column (2)
corresponds to the case where Neptune's obliquity is zero.}
\label{tidesoutcome}
\end{table}

Now that we have given an overview of the tidal evolution, we turn to what the final solutions are when the ensemble of captured
orbits are run through the tidal model.

\subsection{Tidal evolution: Final orbits}
In this subsection we shall present the results of the possible final orbits of Triton after tidal evolution of the captured orbits. We
shall focus on those cases that reach $e=10^{-5}$ within the age of the solar system. Other cases are discarded since they are
incompatible with Triton's current orbit.\\

\begin{figure}
\resizebox{\hsize}{!}{\includegraphics[angle=-90]{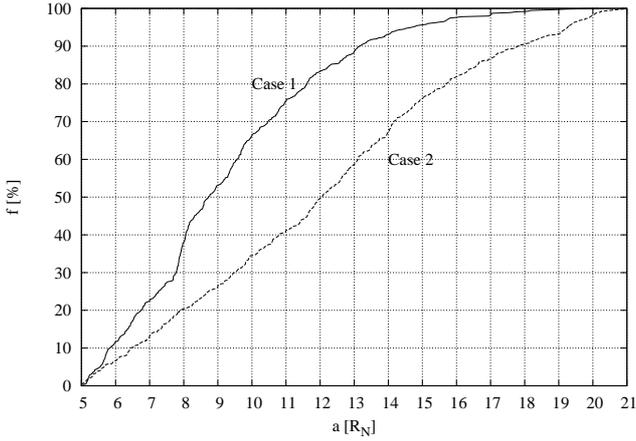}}
\caption{Cumulative distribution of final semi-major axis for objects which reach $e =10^{-5}$ within the age of the solar system. The
solid line is for case 1, the dashed line represents case 2.}
\label{afinal}
\end{figure}

Fig.~\ref{afinal} shows the cumulative distribution in semi-major axis of those objects that reach Triton's current eccentricity
within the age of the solar system for the semi-molten case ($Q_T=10$). The solid line is for case 1, the dashed line represents case
2 (indicated on the plot). There are two important features to note here. First, case 1 yields many orbits with a small final
semi-major axis because of the feature displayed in Fig.~\ref{tideszoom} above: the minimum value of $q$ decreases as $a$ decreases and
the orbits circularise at approximately 2$q_{\rm{min}}$. The second feature is that Triton's current orbit at 14.3~$R_N$ is always in
the upper quartile of the distribution; in the worst case it is in the upper 5\% of the distribution. This result would suggest that if
tides were the dominant mechanism behind circularising Triton's orbit after capture, one would think that Triton should be closer to
Neptune than it is today. \\

\begin{figure}
\resizebox{\hsize}{!}{\includegraphics[angle=-90]{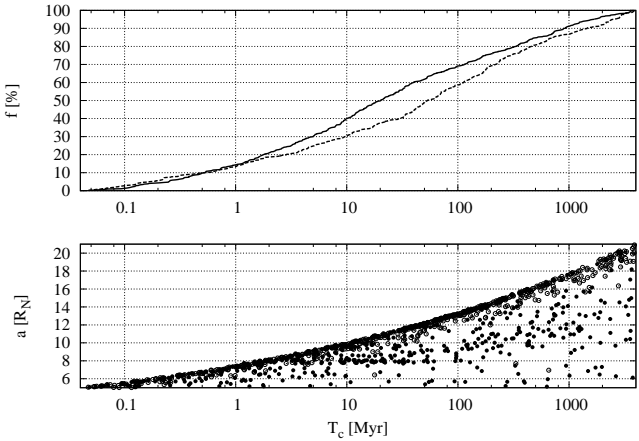}}
\caption{Top panel: Cumulative distribution of time to become circular. The solid line is for case 1, the dashed line represents case
2. Bottom panel: time to become circular vs. final semi-major axis. Bullets present case 1, open circles are case 2. The border between
the filled and open spaces of the plot scales approximately as $a_f \propto T_c^{2/15}$.}
\label{tfinal}
\end{figure}

Fig.~\ref{tfinal} depicts the cumulative distribution of the time it takes for Triton to become circular (top panel) and the final
semi-major axis vs. circularisation time in the bottom panel. Once again in the top panel the solid line presents case 1, the dashed
line is case 2. As one can see, case 1 has a much shorter circularisation time than case 2. This is no surprise because the
circularisation time depends heavily on $q$ as $T_c \propto q^{15/2}$ and since case 1 experiences the lowest values in $q$ it should
circularise the quickest. From the plot one can infer that the median time to become circular ranges from 30~Myr to 70~Myr depending
on the tidal model that is used. However, care has to be taken when using this number because it does not specify the final value of
semi-major axis that is obtained. For that we turn to the bottom panel, which depicts the final semi-major axis vs. the circularisation
time. The bullets present case 1, the open circles are case 2. The line between the filled and open parts of the panel scales as $a_f
\propto T_c^{2/15}$. For Triton at 14~$R_N$, the time to become circular by tides ranges from approximately 200~Myr for most orbits, up
to 4~Gyr for a few exceptional cases, depending on the configuration of the original orbit. Thus, Triton would have reached its current
orbit approximately 200~Myr after its capture. This time scale is consistent with earlier results of McCord (1966), McKinnon
(1984), Goldreich et al. (1989), McKinnon et al (1995) and Correia (2009), who all give a typical time scale of 100-500~Myr for
Triton to become circular, even though not all of the above works use the same tidal model as ours, nor do they take the Kozai
mechanism into account. However, our tidal circularisation time scale is different from that predicted by \'{C}uk \& Gladman
(2005), even if it agrees with other estimates in the literature. We can think of two reasons for this discrepancy. The first is
that \'{C}uk \& Gladman (2005) use a very simplified model for the tides acting on Triton, which does not have the same scaling with
$q$ as our formulation. Their tidal damping time scale proceeds as proportional to $q^6$ while ours goes as $q^{15/2}$. Secondly,
\'{C}uk \& Gladman (2005) average the tidal equation over one Kozai cycle. While from their description it is not clear how this is
done, one would suspect that this results in using an averaged value of $q$ in the tidal equation rather than the minimum value.
The latter controls the time scale for circularisation.\\

After its capture Triton undergoes at least another 3.6~Gyr of tidal evolution due to tides raised on Neptune by Triton. Even though
the influence of tides raised on Neptune by Triton are a couple of orders of magnitude weaker than tides raised on Triton by Neptune,
the subsequent evolution after Triton becomes circular cannot be ignored. For a circular orbit of Triton raising tides on Neptune we
have

\begin{equation}
\dot{a} = \frac{2K_N}{m_ta_T}\Bigl(\frac{\cos i_T \omega_N}{n_T}-1\Bigr).
\end{equation}
Since $\cos i_T\omega_N/n_T \gg 1$, we can ignore the second factor in the brackets, which makes the equation integrable with solution
$a_f = (a_i^{13/2}+Ct)^{2/13}$ where $C = 6k_{2N}m_TR_N^5\Delta t_N \omega_N \cos i_T\sqrt{G/m_N}$. Note that for retrograde
orbits $C<0$ and the orbit shrinks. Over 3.6~Gyr Triton's orbit should have shrunk by about 1.3~$R_N$, from 15.6~$R_N$ to its current
value, making Triton's final position compatible with the tidal model at the 20\% level or less. \\

\begin{figure}
\resizebox{\hsize}{!}{\includegraphics[angle=-90]{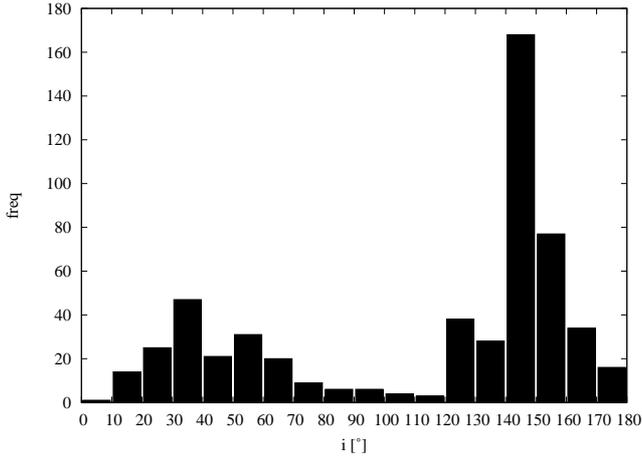}}
\caption{Histograms of the frequency of the final inclination with respect to Neptune's equator for case 1 i.e. with Neptune's
current obliquity.}
\label{ifinal}
\end{figure}

Finally, Fig.~\ref{ifinal} shows histograms of the distribution of the final inclination with respect to Neptune's equator of all
circular orbits that used case 1. There is a large surplus of retrograde objects and the majority of these are situated between
140$^\circ$ and 150$^\circ$, slightly lower than Triton's current value of 157$^\circ$. The large number of objects in this bin can
partially be explained by the following. The inclination of Triton with respect to Neptune's equator ($i$) is related to its
inclination with respect to Neptune's orbit ($I$), Neptune's obliquity ($\varepsilon_N$) and the longitude of the ascending node
($\Omega$) by $\cos i = \cos \varepsilon_N \cos I - \sin \varepsilon_N \sin I \cos \Omega$. Since $\cos i$ is almost a constant for
bodies close to Neptune, the value of $I$ varies between $i+\varepsilon_N$ and $i-\varepsilon_N$ with the circulation time of $\Omega$.
Consequently, this same final value of $i$ can arise from orbits with original inclinations $I+\varepsilon_N$ and $I-\varepsilon_N$,
where the final inclination depends on the original value of $\Omega$. The maximum value of $I$ is 180$^\circ$ and so
$180^\circ-\varepsilon_N = 151.5^\circ$. Thus, the bin between 140$^\circ$ and 150$^\circ$ can sample orbits from the full range
$(I+\varepsilon_N,I-\varepsilon_N)$, but orbits with higher $i$ cannot. Alternatively, orbits with lower values of $i$ sample orbits
with $I <120^\circ$, where the Kozai mechanism operates strongly and can drive orbits to collide with Neptune. These collisions are the
reason for the paucity of orbits with final $i \sim 90^\circ$, even if these orbits are protected from Kozai mechanism at small
semi-major axis because of Neptune's $J_2$. Similarly, the maximum between 30$^\circ$ and 40$^\circ$ can be explained in a similar
manner, though the number difference with the corresponding retrograde case is most likely a result of the retrograde cases being
stable up to larger distances from Neptune. \\

We close this subsection by presenting some statistics for the case where Neptune's obliquity is not zero since this is the most
probable. Of all the Tritons that we analysed, 3\% are captured with an orbit having $q < 1$~$R_N$, so that these collide with Neptune
promptly after their capture. Of the remaining population there are three possible outcomes: circular with final semi-major axis
$a>5$~$R_N$, eccentric orbit with final semi-major axis $a > 5$~$R_N$, and tidal evolution to inside of 5~$R_N$ before becoming
circular. The last category consists of approximately 33\% of the total population, so that some 67\% survive on orbits with final
semi-major axis $a > 5$~$R_N$. This surviving population has only 29\% on orbits with $e > 10^{-5}$ or 19\% of the total population.
Combining these numbers it turns out that a circular Triton with final $a > 5$~$R_N$ occurs approximately 50\% of the time, with
twice as many retrograde vs prograde cases. If Neptune's obliquity were zero, the above statistics are virtually the same.

\subsection{Probabilities}
Next we examine whether or not the existence of Triton is compatible with a capture during the planetary instability. Here we just
concern ourselves
with how many bodies of Triton's size are inferred to have existed from Triton's current orbit. However, before we continue we
should pause for a moment and reflect on what we are interested in. The question is whether or not the capture of a Triton-like body
is plausible, but not be so common that Uranus would have experienced the same event. We can make this argument even stronger:
the Uranian satellite appears to be unperturbed and regular while Neptune's system appears to have been disrupted. For this to occur
it is unimportant if the perturbing body was captured prograde or retrograde, and whether or not the orbit is circular at the current
epoch. Only if the capture is a probable event is it worthy to ask the question about the most likely end states.\\

The probability of Triton's capture through the dissociation of a binary, and its subsequent evolution, is split into the product of
two probabilities: the probability that over its whole lifetime a binary suffers an encounter with Neptune close enough for it to
be disrupted ($P_1$), and the probability that the binary member of Triton's mass is actually captured ($P_2$). Their product is
Triton's capture probability. We could restrict ourselves to only circular and retrograde outcomes, and thus we need to multiply the
above total probability by the probability that after capture Triton reaches a circular orbit at the current epoch and is retrograde
($P_3$). However being circular and retrograde at the current epoch is not necessary for disrupting Neptune's primordial satellite
system. \\

The probability $P_1$ should reflect the encounter history of the binary until its distruption. Each encounter with Neptune softens the
binary and, provided it is not disrupted, after many encounters it has no knowledge of its original binding energy (Parker \&
Kaverlaars, 2010). However, we cannot take this history into account in our current methods but given that the probability of passing
Neptune close enough to be disrupted is approximately $10^{-4}$ and that each planetesimal undergoes approximately 100 encounters, we
believe that our method is accurate enough. It is not necessary to disrupt the binary on its first passage and thus we have used the
total probability of disruption averaged over many encounters rather than just for the first encounter. \\

The value of $P_1$ depends on the distance from Neptune at which the binary gets disrupted, which in turn depends on the mass ratio
and total mass of the binary. Similarly, $P_2$ depends on the system configuration to determine how many Tritons are captured. In
table~\ref{p1} we list the values of $P_1$ and $P_2$ for the binary systems that we considered, in units of percent. The values of
$P_1$ are for encounters that occur during and after the planetary instability, since we have information about how much mass there was
available at this time (Gomes et al., 2005). Since the cumulative distribution of the binary's peri-Neptune distance is linear,
the probability of disruption $P_1 \propto r_{td} \propto (m_1+m_2)^{-1/3}$, which is the approximate trend observed in the table.
\\

\begin{table}
\begin{tabular}{ccccc}
Masses [$m_T$]& $r_{\rm{td}}$ [$R_N$] & $P_1$ [\%] & $P_2$ [\%] & $P_1 \times P_2$ [\%] \\ \hline \\
1.0, 0.1 & 35 & 3.9 & 19.3 & 0.75 \\
1.0, 0.3 & 33 & 3.7 & 19.8 & 0.73 \\
1.0, 1.0 & 29 & 3.1 & 36.9 & 1.1\\
3.0, 1.0 & 23 & 2.1 & 11.8 & 0.25
\end{tabular}
\caption{Table with probabilities $P_1$ and $P_2$ and the total capture probability $P_c = P_1 \times P_2$ for the various binary
systems that we considered.}
\label{p1}
\end{table}
The value of $P_3$ depends on the tidal models that we used. Its values are listed in Table~\ref{probs} and the unit is again
percent. Both cases 1 and 2 are considered. The value of $P=P_1 \times P_2 \times P_3$ is also listed, where we used the average
values of $P_1$ and $P_2$ from the previous table. \\

\begin{table}
\begin{tabular}{ccc}
Case & $P_3$ [\%] & $P$ [\%] \\ \hline \\
1 & 30.1 & 0.21 \\
2 & 29.8 & 0.21
\end {tabular}
\caption{Probabilities $P_3$ and the total combined probability of capture and a circular retrograde orbit at the current epoch,
$P=P_1 \times P_2 \times P_3$, for tidal cases 1 and 2.}
\label{probs}
\end{table}
The total probability of a successful capture, $P_c = P_1 \times P_2$, is on average approximately 1:140. If we then restrict
ourselves to Triton being retrograde and circular at the current epoch, the probability becomes $P$=1:500. We can use these
values to constrain the number of primordial binaries, $N_B$, having one Triton-sized member. The probability of having at least one
Triton capture if we have $N_B$ Triton binaries is 1 minus the probability of its negation, that is, the probability of having no
capture after all $N_B$ cases. Since the events are independent this total probability becomes the product of each individual
probability of not having a capture. Each individual probability of not having a capture is $1-P=139/140$ so that the probability of
having no capture after $N_B$ trials is (139/140)$^{N_B}$. The probability of having at least one capture is of course its complement
$P_{\rm{tot}}=1-(1-P)^N_B = 1-(139/140)^{N_B}$. Since Neptune's system appears to be disrupted and the Uranian system appears to be
regular, the maximum value of $P_{\rm{tot}}$ is approximately 50\%. Solving for $N_B$ we obtain $N_B = 97$ for the number of binaries
at the planetary instability epoch with at least one Triton-sized member. Is this consistent with current theory? Morbidelli et al.
(2009) claim there were $N_T \sim$1\,000 Tritons in the trans-Neptunian disc at the planetary instability epoch, within factor of a
few. This implies a primordial binary population with at least one Triton-sized member, and with the mass ratios that we
considered, of approximately 10\%. The primordial Kuiper Belt population is thought to have consisted of 5\%-40\% binaries (Burns,
2004; Noll et al., 2008; Lin et al., 2010), so that our value of $N_B$ is consistent with this estimate, though only at the lowest
level. However, the above fraction corresponds to smaller size bodies than the ones we are interested in here, and there is some
indirect evidence that the primordial binary population for heavier bodies is lower (Brown et al., 2006). Thus we conclude that the
Agnor \& Hamilton (2006) binary capture scenario during the planetary instability is consistent with Triton's existence. \\

However, in the above argument we did not place any restriction on the final semi-major axis, eccentricity and inclination of Triton.
Placing restraints on the final semi-major axis and inclination will decrease the final probability, $P$, because a fourth probability,
$P_4$, comes into play. It would contain information about the chances of having Triton end up in a specific range of semi-major axis
and/or inclination. This introduces the question of choosing a suitable bin size for the final semi-major axis and/or inclination that
could be compatible with Triton's current orbit. We prefer not to do that here and instead only quote the probability of Neptune ending
up with a circular, retrograde satellite of Triton's mass. \\

Given that Neptune has a large, retrograde satellite and the other giant planets do not, a natural question to ask is why did
this not happen for the other giant planets. The only reason we can think of is that Neptune encounters more planetesimals than the
other giant planets (see table~\ref{encst}). Even though Uranus encounters a similar number, the encounters occur at a greater speed
which decreases the probability of capture. 

\section{Nereid: The fly in the ointment}
In the previous section we have presented the results of our numerical simulations. We concluded that the capture of Triton through
the dissociation of a binary that had a deep encounter with Neptune, followed by subsequent tidal evolution that left Triton
semi-molten (McKinnon, 1984), is enough to place Triton on its current circular, retrograde orbit. No extra ingredients, such as the
collision with a hypothetical regular satellite of Neptune (Goldreich et al., 1989), the presence of a gas disc (McKinnon \& Leith,
1995) or a debris disc left over from the mutual collisions among the members of a fictitious regular satellite system (\'{C}uk \&
Gladman, 2005) are needed (Correia, 2009). Does that mean that we are done? \\

\begin{figure}
\resizebox{\hsize}{!}{\includegraphics[angle=-90]{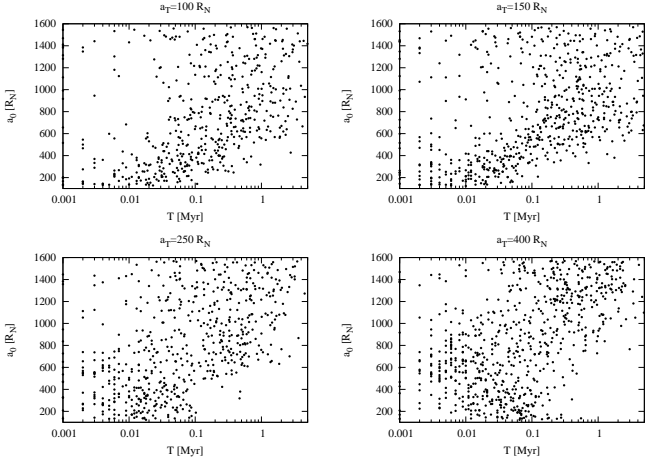}}
\caption{Lifetime vs. initial semi-major axis of a fictitious swarm of irregular satellites of Neptune that were perturbed by Triton.
The initial pericentre of Triton was set at 7~$R_N$ but the semi-major axis was varied (see titles above panels for values).}
\label{tlost}
\end{figure}

\begin{figure}
\resizebox{\hsize}{!}{\includegraphics[angle=-90]{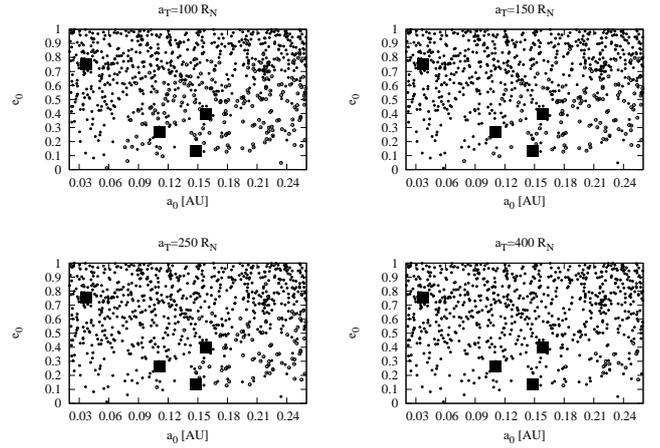}}
\caption{Original semi-major axis vs. initial eccentricity of a fictitious swarm of irregular satellites of Neptune that were
perturbed by Triton. Bullets show objects that are lost, while the open circles show objects that survive. Large filled squares show
the positions of Neptune's currently-known irregular satellites Nereid, Halimede, Sao and Laomedeia. The initial pericentre of Triton
was set at 7~$R_N$ but the semi-major axis was varied (see titles above panels for values).}
\label{lost}
\end{figure}

No, because there are several issues that we did not address. The first of these is what would have happened to Neptune's irregular
satellites if Triton were captured during the planetary instability, when the other irregular satellites were captured too
(Nesvorn\'{y} et al., 2007). While Triton remained on a highly-eccentric orbit, with a semi-major axis comparable to that of the other
irregular satellites, it greatly perturbs the rest of this population (\'{C}uk \& Gladman, 2005). Nereid in particular is difficult to
keep, and \'{C}uk \& Gladman (2005) conclude that it will be lost within 0.1~Myr. However, the initial conditions of \'{C}uk \& Gladman
(2005) are somewhat artificial and direct comparison with our post-capture orbits is difficult. Therefore we have performed similar
simulations, in which we place Triton on an eccentric orbit with semi-major axis ranging from 100 to 400 Neptune radii, pericentre
distance $q \sim 7$~$R_N$ and investigated how its presence affected Nereid and a swarm of other irregular satellites. We placed 1\,000
fictitious irregular satellites around Neptune where the initial conditions were taken from Nesvorn\'{y} et al. (2007). The Sun was
added as an external perturber. It turned out that Nereid is always lost through collision with Neptune within 0.1~Myr, even if it was
captured with an initially almost circular orbit. The longest stability was found when Triton's semi-major axis was less than
100~$R_N$. Using the tidal model, Triton needs of the order of $\sim$100~Myr to become circular. The eccentric phase lasts for
approximately $T_a =a/\dot{a}$ which for highly-eccentric orbits becomes

\begin{equation}
T_a = 1.8 \Bigl(\frac{q}{7\,R_N}\Bigr)^{15/2}\Bigl(\frac{a}{100\,R_N}\Bigr)^{1/2}\quad {\rm{Myr}}.
\end{equation}
Thus when Triton's semi-major axis is 200~$R_N$ and $q \sim 7$~$R_N$ it stays eccentric for approximately 3 Myr, but if
$a=100$~$R_N$ it is not even 2 Myr. This simple approximation does not take Kozai mechanism into account, which increases the time by
approximately one order of magnitude. Thus, with this model Triton stays eccentric for far too long and we lose Nereid. We have
presented the results of some numerical simulations in Fig.~\ref{tlost}. The panels plot the time a satellite is lost vs.
its initial semi-major axis. As can be seen, even when Triton has an orbit with just $a=100$~$R_N$, Nereid (at $a=222$~$R_N$) is lost
within 0.1~Myr, even when Nereid's initial eccentricity is close to zero.  Finally, Fig.~\ref{lost} shows the original semi-major
axis and eccentricity of the irregular satellites that are lost (bullets) and that survive (open circles). The big filled squares
indicate Neptune's current irregular satellites Nereid, Halimede, Sao and Laomedeia. The farthest two, Psamathe and Neso, are off the
scale to the right. As one can see, Nereid is always lost, even if it was captured with a low eccentricity. Only satellites captured
with a low eccentricity and which also have a semi-major axis $a \gtrsim 2a_T$ survive.\\

\begin{figure}
\resizebox{\hsize}{!}{\includegraphics[angle=-90]{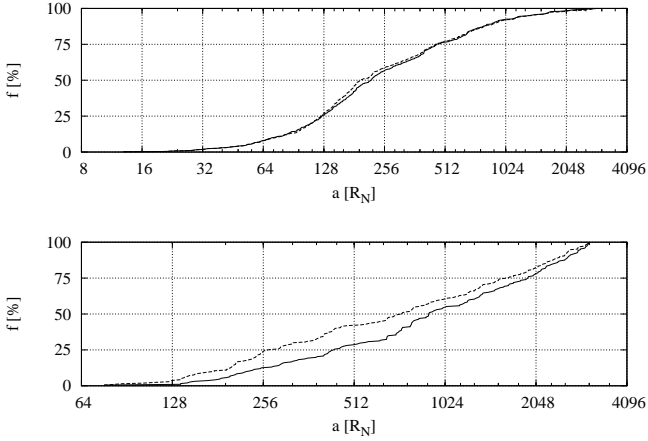}}
\caption{Top panel: Cumulative distribution of semi-major axis upon capture of objects that reach eccentricity $e=10^{-5}$ within the
age of the solar system. The solid line represents case 1, the dashed line is case 2. Bottom panel: the same as the top panel but for
orbits with final $a > 5$~$R_N$ and $e > 10^{-5}$.}
\label{cuma0}
\end{figure}

There are five ways out of this dilemma that we can think of. The first is that Triton could have been captured with a very
small semi-major axis, smaller than $\sim 50$~$R_N$. In the top panel of Fig.~\ref{cuma0} we plotted the cumulative distribution of the
original semi-major axis upon capture for all objects which reached an eccentricity $e = 10^{-5}$ within the age of the solar system.
The solid line is for case 1, the dashed line represents case 2. As one can see, only 5\% of the time is Triton captured with a
semi-major axis $\lesssim 50$~$R_N$. The median value is around 200~$R_N$, approximately where Nereid is ($a = 222$~$R_N$), which is
much lower than the median value upon capture (Fig.~\ref{captcum}). Thus, capture at small semi-major axis is unlikely, but not
impossible. For reference, the bottom panel contains the cumulative semi-major axis distribution of objects with $a > 5$~$R_N$ and $e >
10^{-5}$. Most of these are still exhibiting Kozai oscillations and have undergone very little tidal evolution.\\

A second possible exit strategy is to argue that Nereid was not captured by the mechanism of Nesvorn\'{y} et al. (2007) and
instead was a regular satellite of Neptune that was scattered outwards by Triton. We performed a series of numerical simulations to
test this hypothesis. We placed a large number of test particles on circular, equatorial orbits around Neptune with a maximum
semi-major axis of 45~$R_N$, which is twice as far as Oberon is from Uranus. We have not witnessed any regular satellite being
scattered to Nereid's orbit by Triton. Instead, they all collide with Neptune within a few tens of thousands of years, or they are
ejected by Triton when their orbits reach $q \lesssim 3$~$R_N$ and their orbital angular momentum is at a minimum. Even if Triton
could place a regular satellite on Nereid's orbit, we are still faced with the dilemma of keeping it there afterwards. In addition, we
have one additional argument against the scattering scenario. \\

The measured rotation period of Nereid is 13.6~h (Grav et al., 2003). This has two implications: i) either Nereid was captured on its
current orbit and thus the rotation period that we see is a remnant from when it formed, because it is too far away from Neptune to be
tidally despun, or ii) it used to be on a 13.6~h orbit around Neptune and was scattered to its current orbit by Triton. A period of
13.6~h corresponds to a semi-major axis of approximately 3~$R_N$, closer in than Proteus at 4.7~$R_N$. If Nereid was any further from
Neptune its rotation period would be longer than it is now. So it is extremely unlikely that Triton scattered a synchonous Nereid from
close to Neptune to its current orbit without messing up the rest of the system, in particular the satellites Proteus and Galatea. Thus
a captured origin for Nereid is the most plausible. If this capture occured around the same time as that of Triton, Nereid would
be lost through collision with Neptune. \\

A third scenario has been proposed by Desch \& Porter (2010), who investigated the idea of Triton having been a satellite of a 2
Earth-mass fictitious planet called Amphitrite. When the binary Amphitrite-Triton suffered a close approach with Neptune, Triton was
captured in orbit around Neptune. Desch \& Porter (2010) speculated that Amphitrite could later collide with Uranus to produce its
axial tilt or with Neptune itself to account for its excess heat radiation. Assuming that Amphitrite collided with Neptune and
that this caused the capture of Triton, Desch \& Porter (2010) state the probability of capturing Triton in this manner is approxiately
20\%-40\% for encounter velocities at Neptune's Hill sphere of less than 3~km~s$^{-1}$. Their typical pericentre distance of
Triton after captre is 7~$R_N$, for original Amphitrite-Triton semi-major axis shorter than 40~$R_N$. Unfortunately Desch \& Porter
(2010) do not give any information about the typical semi-major axis of Triton after capture, so that it is unclear if this capture
mechanism is able to capture Triton at a short-enough semi-major axis in order to prevent the loss of Nereid. In addition, for low
encounter velocities and long initial semi-major axis of the binary, the pericentre distance of Triton upon capture is large and tides
may not be able to circularise it within the age of the Solar System.\\

A fourth scenario involves the debris disc proposed by \'{C}uk \& Gladman (2005). However, as we stated in the introduction, it is
not clear whether or not Neptune's hypothetical regular satellites will grind themselves down to a debris disc before one of them
collides with Triton and shatters both Triton and itself. In principle the satellites collide with each other because Triton forces
their eccentricities. This forcing is inversely proportional to the semi-major axis ratio of the regular satellites and Triton, and
Triton's eccentricity. Furthermore, if Triton's semi-major axis is very long it is energetically more favourable for the satellites to
eject Triton. Thus we envision three possible outcomes as a function of increasing original semi-major axis of Triton after capture.
When Triton's semi-major axis is short, mutual collision among the satellites, which form the debris disc, is the most likely outcome
because Triton induces a large eccentricity in the satellites. For intermediate semi-major axis of Triton, the forced eccentricities of
the satellites are not large enough for them to cross each other and Triton will hit one of these satellites before it is ejected. For
very long semi-major axes of Triton, it will be ejected before a collision occurs. We reserve investigating this scenario for future
work.\\

\begin{figure}
\resizebox{\hsize}{!}{\includegraphics[angle=-90]{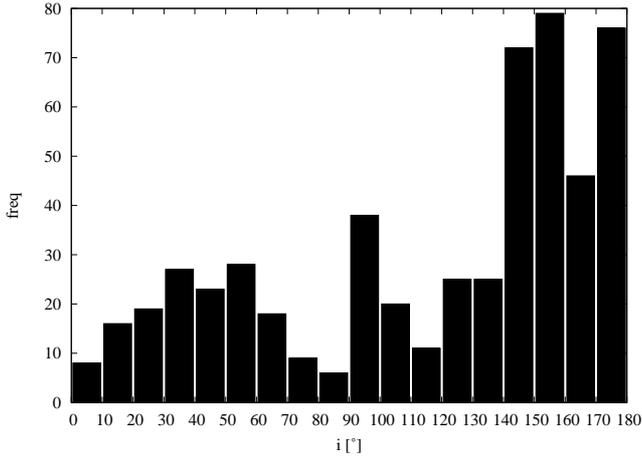}}
\caption{Histograms of the frequency of the final inclination with respect to Neptune's equator for case 2 i.e. without Neptune's
obliquity.}
\label{ifinal2}
\end{figure}

The fifth, most plausible, scenario is that Triton was captured and circularised before the planetary instability, something which was
already suggested
by Vokrouhlick\'{y} et al. (2009) but for different reasons. This early capture scenario would solve the problem of destabilising
Neptune's other irregular satellites, such as Nereid, because it was unlikely that they were already there. Since Neptune
might not have had a circumplanetary gas disc but just a cooling, rotating envelope (Ayliffe \& Bate, 2009), the perturbations from
Triton could have disturbed this system enough to prevent the formation of a regular satellite system such as that of Uranus, while
decreasing its orbital energy at the same time. Proteus' current position would be indicative of the minimum distance Triton reached
after its capture prior to reaching its current orbit. In addition, at this early stage Neptune's obliquity may have been close
to 0, increasing the likelihood of Triton ending up on its current orbit (see Fig.~\ref{afinal}). Neptune's low obliquity would also
have changed the final inclination distribution, which we have plotted in Fig.~\ref{ifinal2}. As one can see, the retrograde orbits are
more evenly spread over a larger interval. We should add that it is possible that the final semi-major axis and inclination
distribution of captured objects before the planetary instability would be different than that presented above, because the binary
encounters with Neptune would have occurred at different velocities since Neptune was most likely closer to the Sun, and because it was
not migrating. However, we do not think that the final results would be qualitatively very different from what has been presented
above, although the probabilities discussed earlier would most likely change. The early capture scenario also revokes the need
for a debris disc resulting from mutual collisions among fictitious regular satellites of Neptune (\'{C}uk \& Gladman, 2005) or a
collision with such a fictitious satellite (Goldreich et al, 1989) and only requires the minimum ingredients: a binary capture and
tidal evolution.\\

\section{Summary and conclusions}
We have made an attempt at determining the origin of Triton, Neptune's large, circular, retrograde satellite. Our work is based on
two assumptions. The first is that Triton was captured through a binary exchange (Agnor \& Hamilton, 2006) and that the subsequent
shrinking and circularisation of the orbit occurred through tides only (Correia, 2009). We performed numerical simulations of the
migration of Neptune in the framework of the Nice model (Tsiganis et al., 2005) and recorded the close encounters that the
planetesimals suffered with Neptune. The closest of these encounters were re-enacted with binaries of various mass ratios to determine
whether the capture of Triton would occur, and for what parameters. The re-enaction experiments yielded a set of initial orbits for
Triton just after its capture. These post-capture orbits were integrated using the tidal model of Mignard (1979, 1980) and
Hut (1981), as was done by Correia (2009). However, we added the effects of Kozai mechanism induced by the Sun (Kozai, 1962) and the
effect of Neptune's figure (e.g. Kinoshita \& Nakai, 1991). The former will cause oscillations in Triton's eccentricity and
inclination. Thus Triton's orbital angular momentum, which is constant when only tides are taken into account, is no longer
conserved. The perturbations from the figure of Neptune will overrule the Kozai effect once Triton is close enough to Neptune, so that
the eccentricity and inclination oscillations cease. We integrated the tidal equations until Triton's orbit was circular ($e <
10^{-5}$).\\

We find that a binary capture and tidal evolution are sufficient to reproduce Triton's current orbit, even though the tidal model
would predict Triton to be closer to Neptune than its current position at 14.3~$R_N$. The probability of Neptune having a Triton-mass
satellite is 0.7\%. From this we deduce there were $\sim 100$ binaries in the trans-Neptunian disc with at least one Triton-sized
member. This number is consistent with theoretical predictions, though at the low end. \\

The typical time for a final orbit at 14~$R_N$ to become circular is of the order of 200~Myr, much shorter than the age of the solar
system or the time between the formation of the giant planets and the planetary instability. However, the Kozai oscillations caused by
the perturbations from the Sun increase the time that Triton stays in an eccentric orbit. The perturbations from an eccentric Triton
destabilise Nereid on a time scale of less than 0.1~Myr, so that its existence is in contradiction with our model. The most plausible
exit strategies are either capture of Triton at small semi-major axis (a rare event), or a capture before the planetary instability
when Nereid was not there. The early capture of Triton would remove the need for a fast circularisation of its orbit, decrease the
possibility of a collision with an existing regular satellite and also increase the probability of Triton ending up on its current
orbit because Neptune's obliquity may have been close to zero. Thus, we suggest that Triton was captured shortly after Neptune's
formation through a binary encounter and was circularised to its current orbit through tides, while possibly disturbing Neptune's
rotating envelope (Ayliffe \& Bate, 2009) and preventing the formation of other regular satellites.\\

{\footnotesize {\bf Acknowledgements} \\
The authors wish to express their gratitude to Alexandre Correia for making available to us some
of his data and for many helpful discussions. RB sincerely thanks Germany's Helmholtz Alliance through their 'Planetary Evolution and
Life' programme for financial support. RG and EN express their gratitude towards Brazil's CNPq and FAPERJ for their funding. Part of
this work was accomplished while RB visited ON/MCT and he thanks his hosts RG and EN for their hospitality and Brazil's CNPq for
financing his trip.} 
\section{Bibliography}
{\footnotesize
Agnor, C., Asphaug, E.\ 2004.\ Accretion Efficiency during Planetary Collisions.\ The Astrophysical Journal 613, L157-L160.\\
Agnor, C.~B., Hamilton, D.~P.\ 2006.\ Neptune's capture of its moon Triton in a binary-planet gravitational encounter.\ Nature 441,
192-194. \\
Anderson, J.~D., Jacobson, R.~A., Lau, E.~L., Moore, W.~B., Schubert, G.\ 2001.\ Io's gravity field and interior structure.\ Journal of
Geophysical Research 106, 32963-32970. \\
Ayliffe, B.~A., Bate, M.~R.\ 2009.\ Circumplanetary disc properties obtained from radiation hydrodynamical simulations of gas accretion
by protoplanets.\ Monthly Notices of the Royal Astronomical Society 397, 657-665. \\
Benner, L.~A.~M., McKinnon, W.~B.\ 1995.\ Orbital behavior of captured satellites: The effect of solar gravity on Triton's postcapture
orbit..\ Icarus 114, 1-20. \\
Bernstein, G.~M., Trilling, D.~E., Allen, R.~L., Brown, M.~E., Holman, M., Malhotra, R.\ 2004.\ The Size Distribution of
Trans-Neptunian Bodies.\ The Astronomical Journal 128, 1364-1390. \\
Brown, M.~E., van Dam, M. A., Bouchez, A. H., Le Mignant, D., Campbell, R. D., Chin, J. C. Y., Conrad, A., Hartman, S. K., Johansson,
E. M., Lafon, R. E., Rabinowitz, D. L., Stomski, P. J., Jr., Summers, D. M., Trujillo, C. A., Wizinowich, P. L. 2006.\ Satellites of
the Largest Kuiper Belt Objects.\ The Astrophysical Journal 639, L43-L46. \\
Bulirsch, R., Stoer, J. 1966. \ Numerical treatment of ordinary differential equations by extrapolation methods. \ Numerische
Mathematik 8, 1-13. \\
Burns, J.~A.\ 2004.\ Planetary science:  Double trouble.\ Nature 427, 494-495.\\
Canup, R.~M., Ward, W.~R.\ 2006.\ A common mass scaling for satellite systems of gaseous planets.\ Nature 441, 834-839. \\
Chambers, J.~E.\ 1999.\ A hybrid symplectic integrator that permits close encounters between massive 
bodies.\ Monthly Notices of the Royal Astronomical Society 304, 793-799. \\
Correia, A.~C.~M.\ 2009.\ Secular Evolution of a Satellite by Tidal Effect: Application to Triton.\ The Astrophysical Journal 704,
L1-L4. \\
{\'C}uk, M., Gladman, B.~J.\ 2005.\ Constraints on the Orbital Evolution of Triton.\ The Astrophysical Journal 626, L113-L116. \\
Desch, S., Porter, S.\ 2010.\ Amphitrite: A Twist on Triton's Capture.\ Lunar and Planetary Institute Science Conference Abstracts 41,
2625. \\
Duncan, M., Quinn, T., Tremaine, S.\ 1987.\ The formation and extent of the solar system comet cloud.\ The Astronomical Journal 94,
1330-1338. \\
Efroimsky, M., Lainey, V.\ 2007.\ Physics of bodily tides in terrestrial planets and the appropriate scales of dynamical evolution.\
Journal of Geophysical Research (Planets) 112, 12003. \\
Fern\'{a}ndez, J.~A.\ 1981.\ New and evolved comets in the solar system.\ Astronomy and Astrophysics 96, 26-35. \\
Fern\'{a}ndez, J.~A.\ 1997.\ The Formation of the Oort Cloud and the Primitive Galactic Environment.\ Icarus 129, 106-119. \\
Goldreich, P., Murray, N., Longaretti, P.~Y., Banfield, D.\ 1989.\ Neptune's story.\ Science 245, 500-504. \\
Gomes, R.~S., Morbidelli, A., Levison, H.~F.\ 2004.\ Planetary migration in a planetesimal disk: why 
did Neptune stop at 30 AU?.\ Icarus 170, 492-507. \\
Gomes, R., Levison, H.~F., Tsiganis, K., Morbidelli, A.\ 2005.\ Origin of the cataclysmic Late Heavy Bombardment period of the
terrestrial planets.\ Nature 435, 466-469. \\
Grav, T., Holman, M.~J., Kavelaars, J.~J.\ 2003.\ The Short Rotation Period of Nereid.\ The Astrophysical Journal 591, L71-L74. \\
Hamilton, D.~P., Krivov, A.~V.\ 1997.\ Dynamics of Distant Moons of Asteroids.\ Icarus 128, 241-249. \\
Hut, P.\ 1981.\ Tidal evolution in close binary systems.\ Astronomy and Astrophysics 99, 126-140. \\
Jacobson, R.~A., Riedel, J.~E., Taylor, A.~H.\ 1991.\ The orbits of Triton and Nereid from spacecraft and earthbased observations.\
Astronomy and Astrophysics 247, 565-575. \\
Jacobson, R.~A., Campbell, J.~K., Taylor, A.~H., Synnott, S.~P.\ 1992.\ The masses of Uranus and its major satellites from Voyager
tracking data and earth-based Uranian satellite data.\ The Astronomical Journal 103, 2068-2078.\\
Karato, S.-I. 2008. Deformation of Earth Materials: An Introduction to the Rheology of Solid Earth. Cambridge Univ. Press,
Cambridge, UK \\
Kinoshita, H., Nakai, H.\ 1991.\ Secular perturbations of fictitious satellites of Uranus.\ Celestial Mechanics and Dynamical Astronomy
52, 293-303. \\
Kinoshita, H., Nakai, H.\ 2007.\ General solution of the Kozai mechanism.\ Celestial Mechanics and Dynamical Astronomy 98, 67-74. \\
Kozai, Y.\ 1962. \ Secular perturbations of asteroids with high inclination and eccentricity. The Astronomical Journal 67, 591--598. \\
Lainey, V., Dehant, V., P{\"a}tzold, M.\ 2007.\ First numerical ephemerides of the Martian moons.\ Astronomy and Astrophysics 465,
1075-1084. \\
Lainey, V., Arlot, J.-E., Karatekin, {\"O}., van Hoolst, T.\ 2009.\ Strong tidal dissipation in Io and Jupiter from astrometric
observations.\ Nature 459, 957-959.\\
Levison, H.~F., Morbidelli, A., Vanlaerhoven, C., Gomes, R., Tsiganis, K.\ 2008.\ Origin of the structure of the Kuiper belt during a
dynamical instability in the orbits of Uranus and Neptune.\ Icarus 196, 258-273. \\
Levison, H.~F., Bottke, W.~F., Gounelle, M., Morbidelli, A., Nesvorn{\'y}, D., Tsiganis, K.\ 2009.\ Contamination of the asteroid belt
by primordial trans-Neptunian objects.\ Nature 460, 364-366. \\
Lin, H.-W., Kavelaars, J.~J., Ip, W.-H., Gladman, B.~J., Petit, J.~M., Jones, R.~L., Parker, J.~W.\ 2010.\ On the Detection of Two New
Trans-Neptunian Binaries from the CFEPS Kuiper Belt Survey.\ Publications of the Astronomical Society of the Pacific 122, 1030-1034. \\
MacDonald, G.~J.~F.\ 1964.\ Tidal Friction.\ Reviews of Geophysics and Space Physics 2, 467-541.\\
Marty, J.~C., Balmino, G., Duron, J., Rosenblatt, P., Le Maistre, S., Rivoldini, A., Dehant, V., van Hoolst, T.\ 2009.\ Martian gravity
field model and its time variations from MGS and Odyssey data.\ Planetary and Space Science 57, 350-363. \\
McCord, T.~B.\ 1966.\ Dynamical evolution of the Neptunian system.\ The Astronomical Journal 71, 585.\\
McKinnon, W.~B.\ 1984.\ On the origin of Triton and Pluto.\ Nature 311, 355-358. \\
McKinnon, W.~B., Leith, A.~C.\ 1995.\ Gas drag and the orbital evolution of a captured Triton..\ Icarus 118, 392-413. \\
Mignard, F.\ 1979.\ The evolution of the lunar orbit revisited. I.\ Moon and Planets 20, 301-315.\\
Mignard, F.\ 1980.\ The evolution of the lunar orbit revisited. II.\ Moon and Planets 23, 185-201. \\
Morbidelli, A., Levison, H.~F., Tsiganis, K., Gomes, R.\ 2005.\ Chaotic capture of Jupiter's Trojan asteroids in the early Solar
System.\ Nature 435, 462-465. \\
Morbidelli, A., Tsiganis, K., Crida, A., Levison, H.~F., Gomes, R.\ 2007.\ Dynamics of the Giant Planets of the Solar System in the
Gaseous Protoplanetary Disk and Their Relationship to the Current Orbital Architecture.\ The Astronomical Journal 134, 1790-1798. \\
Morbidelli, A., Levison, H.~F., Bottke, W.~F., Dones, L., Nesvorn{\'y}, D.\ 2009.\ Considerations on the magnitude distributions of
the Kuiper belt and of the Jupiter Trojans.\ Icarus 202, 310-315. \\
Nesvorn{\'y}, D., Vokrouhlick{\'y}, D., Morbidelli, A.\ 2007.\ Capture of Irregular Satellites during Planetary Encounters.\ The
Astronomical Journal 133, 1962-1976. \\
Noll, K.~S., Grundy, W.~M., Chiang, E.~I., Margot, J.-L., Kern, S.~D.\ 2008.\ Binaries in the Kuiper Belt.\ The Solar System Beyond
Neptune. 345-363. M. A. Barucchi, H. Boehnhardt, D. P. Cruikshank, A. Morbidelli (eds). University of Arizona Press, Tucson, AZ,
USA.\\ 
\"{O}pik, E. \ 1976. \ Interplanetary Encounters. \ Elsevier Scientific Publishing, Amsterdam. \\
Parker, A.~H., Kavelaars, J.~J.\ 2010.\ Destruction of Binary Minor Planets During Neptune Scattering.\ The Astrophysical Journal 722,
L204-L208. \\
Tsiganis, K., Gomes, R., Morbidelli, A., Levison, H.~F.\ 2005.\ Origin of the orbital architecture of the giant planets of the Solar
System.\ Nature 435, 459-461. \\
Vokrouhlick{\'y}, D., Nesvorn{\'y}, D., Levison, H.~F.\ 2008.\ Irregular Satellite Capture by Exchange Reactions.\ The Astronomical
Journal 136, 1463-1476. \\
Ward, W.~R., Canup, R.~M.\ 2003.\ Viscous Evolution of an Impact Generated Water/Rock Disk Around Uranus.\ Bulletin of the American
Astronomical Society 35, 1046. \\
Williams, J. G., Boggs, D. H., Ratcliff, J. T.\ 2005. Lunar Fluid Core and Solid Body Tides, LPI 36, 1503.\\
Yoder, C.~F., Konopliv, A.~S., Yuan, D.~N., Standish, E.~M., Folkner, W.~M.\ 2003.\ Fluid Core Size of Mars from Detection of the Solar
Tide.\ Science 300, 299-303. \\
}
\end{document}